\documentclass[twocolumn,epjc3]{svjour3}          % twocolumn

\usepackage[T1]{fontenc}

\usepackage{graphicx}
\usepackage{mathptmx}      % use Times fonts if available on your TeX system
\usepackage{flushend}
\usepackage[numbers,sort&compress]{natbib}
\usepackage[colorlinks,citecolor=blue,urlcolor=blue,linkcolor=blue]{hyperref}
\journalname{Eur. Phys. J. C}

\usepackage[italic]{hepnames}

\usepackage{breqn}             % These four lines are
\makeatletter                  % used to line-break equations
\let\cat@comma@active\@empty   % automatically
\makeatother                   % use \begin{dmath} ... \end{dmath} for eqns

\begin{document}

\title{Charmonium spectroscopy motivated by general features of pNRQCD}

\author{Raghav Chaturvedi \and Ajay Kumar Rai}

\institute{Department of Applied Physics, Sardar Vallabhbhai National
Institute of Technology, Surat 395007, Gujarat, {\it INDIA}}

\date{Received: date / Accepted: date}
\maketitle

\begin{abstract}
Mass spectrum of charmonium is computed in the framework of potential non-relativistic quantum chromodynamics. \textit{O}$(1/m)$ and \textit{O}$(1/m^2)$ relativistic corrections to the Cornell potential and spin-dependent potential have been added, and is solved numerically. New experimentally observed and modified positive and negative parity states like ${{\boldsymbol \psi}{(4230)}}$, ${{\boldsymbol \psi}{(4260)}}$, ${{\boldsymbol \psi}{(4360)}}$, ${{\boldsymbol \psi}{(4390)}}$, ${{\boldsymbol \psi}{(4660)}}$, ${{\boldsymbol \chi}_{{c1}}{(4140)}}$ and ${{\boldsymbol \chi}_{{c1}}{(4274)}}$ near open-flavor threshold have also been studied. We explain them as admixtures of S-D wave states and P-wave states.
Apart from these states, some other states like ${{\mathit X}{(3915)}}$, ${{\boldsymbol \chi}_{{c1}}{(3872)}}$, ${{\boldsymbol \psi}{(3770)}}$ and ${{\boldsymbol \psi}{(4160)}}$ have been identified as $2^3P_0$, $2^3P_1$, $1^3D_1$ and $2^3D_1$ states. Subsequently, the electromagnetic transition widths and $\gamma\gamma$, $e^+e^-$, light hadron and $\gamma\gamma\gamma$ decay widths of several states are calculated at various leading orders. All the calculated results are compared with experimental and results from various theoretical models.
\end{abstract}

\section{Introduction}
\label{intro}

Remarkable experimental progress has been made in recent years in the field of heavy flavour hadrons specially charmonia. All the narrow charmonium states below open charm threshold($D\bar{D}$) have been observed experimentally and they have been successfully studied theoretically by many approaches like lattice QCD \cite{McNeile:2012qf}, chiral perturbation theory \cite{Gasser:1983yg}, heavy quark effective field theory \cite{Neubert:1993mb}, QCD sum rules \cite{Veliev:2010vd}, NRQCD \cite{Bodwin:1994jh}, dynamical equations based approaches like\\ Schwinger-Dyson and Bethe-Salpeter equations(BSE) \cite{LlewellynSmith:1969az,Mitra:2001ns,Ricken:2000kf,Mitra:1990av} and some potential models \cite{Vinodkumar:1999da,Pandya:2001hx,KumarRai:2005yr,Rai:2006bt,Rai:2006wm,Pandya:2008vf,Vinodkumar:2008zd,Rai:2008zza,Pandya:2014qma}. However, there are many questions related to charmonium physics in the region above $D\bar{D}$ threshold. $X$ and $Y$ states above $D\bar{D}$ threshold have been reported with unusual properties which are yet to be explained completely. These states however could be exotic states, mesonic molecules multi quark states or even admixtures of lower lying charmonia states which have been broadly put forward in \cite{Brambilla:2010cs} and references therein.
Charmoniumlike states having normal quantum numbers have similar masses when compared to normal charmonium, thus in order to study and understand the nature of higher mass states near $D\bar{D}$ threshold it is necessary to have better understanding of lower lying charmonium states.\\
Formerly ${{\mathit X}{(3872)}}$ (Now, ${{\boldsymbol \chi}_{{c1}}{(3872)}}$) was studied for the first time at Belle \cite{Choi:2003ue} in $2003$ in exclusive decay of ${{\mathit B}^{\pm}}$ which was later produced by ${{\mathit p}^{+}}{{\mathit p}^{-}}$ collision \cite{Acosta:2003zx} as well. Successively ${{\mathit X}{(3872)}}$ was studied theoretical as exotic state by \cite{Braaten:2018eov,Kalashnikova_2019}, as pure chamonium state by \cite{Swanson:2004pp,Hanhart:2007yq,Aceti:2012cb}, as meson-meson molecular structure by \cite{Swanson:2004pp,Hanhart:2007yq,Aceti:2012cb}, as tetra-quark state and as charmonia core plus higher fock components due to coupling to meson meson continuum by \cite{Ferretti:2013faa,Barnes:2005pb,Pennington:2007xr,Danilkin:2010cc,Karliner:2014lta,Cardoso:2014xda,Badalian:2015dha}. Recently in PDG \cite{Tanabashi:2018oca} ${{\mathit X}{(3872)}}$ was renamed ${{\boldsymbol \chi}_{{c1}}{(3872)}}$ considering it as a potential charmonium candidate and theoretically supported by \cite{Kher:2018wtv} as pure charmonium candidate. Also CDF collaboration \cite{Abulencia:2006ma} explained ${{\mathit X}{(3872)}}$ particle as a conventional charmonium state with J$^{PC}$ $1^{++}$ or $2^{-+}$.\\
${{\boldsymbol \chi}_{{c1}}{(4140)}}$ previously known as ${{\boldsymbol X}{(4140)}}$ or ${{\boldsymbol Y}{(4140)}}$  discovered by CDF \cite{Aaltonen:2009tz,Aaltonen:2011at} in $2008$ near ${\boldsymbol J / \psi \phi}$ threshold, was later confirmed by D0 and CMS \cite{Abazov:2013xda,Chatrchyan:2013dma}. The result of the state was negative in B decays at Belle \cite{Brodzicka:2010zz,Shen:2009vs}, LHCb \cite{Aaij:2012pz} and BABAR \cite{Lees:2014lra}. The CDF collaboration in $2011$ observed ${{\mathit X}{(4140)}}$ with statistical significance greater than 5 standard deviations and also found evidence for another state ${{\boldsymbol X}{(4274)}}$ now known as ${{\boldsymbol \chi}_{{c1}}{(4274)}}$ with mass $4274.4{}^{+8.4}_{-6.7}\pm1.9$ MeV \cite{Aaltonen:2011at}. LHCb in $2017$ confirmed ${{\boldsymbol \chi}_{{c1}}{(4140)}}$ and ${{\boldsymbol \chi}_{{c1}}{(4274)}}$ with masses $4146.5\pm4.5{}^{+4.6}_{-2.8}$ and $4273.3\pm8.3{}^{+17.2}_{-3.6}$ MeV respectively. Both having $J^{PC}=1^{- -}$ as reported by \cite{Tanabashi:2018oca}. ${{\boldsymbol X}{(4140)}}$ has been studied by many theoretical studies as a molecular state, tetra-quark state or a hybrid state \cite{Liu:2009ei,Branz:2009yt,Albuquerque:2009ak,Ding:2009vd,Stancu:2009ka,Wang:2015pea,Anisovich:2015caa,Wang:2009ue,Mahajan:2009pj}. \cite{Lu:2016cwr} has suggested ${{\boldsymbol X}{(4274)}}$ to be ${\boldsymbol \chi}_{c1}3^3P_1$ state,  \cite{Bhavsar:2018umj} has studied ${{\boldsymbol \chi}_{{c1}}{(4140)}}$ as an admixture of P wave states.\\
${{\mathit X}{(3915)}}$ formerly ${{\boldsymbol \chi}_{{c0}}{(3915)}}$ was observed by Belle collaboration \cite{Uehara:2009tx} with mass $3915\pm 3 \pm 2$ MeV in photon-photon collision and experimental analysis \cite{Tanabashi:2018oca} presented $\mathit J{}^{PC} = 2{}^{++}$ for the same. This let to its assignment as $2^3P_0$ by \cite{Liu:2009fe} and BABAR \& SLAC \cite{Lees:2012xs}, and $2^3P_2$ by \cite{Branz:2010rj}.\\
${{\boldsymbol \psi}{(4230)}}$ previously known as ${{\mathit X}{(4230)}}$ first observed at BESIII \cite{Ablikim:2014qwy} in $2015$ as one of the two resonant structure in ${{\mathit e}^{+}}{{\mathit e}^{-}}\rightarrow{{\boldsymbol \omega}}{{\boldsymbol \chi}_{{c0}}}$ with statistical significance more than $9\sigma$. Having $J^{PC}=1^{- -}$ this state shows properties different from a conventional ${{\mathit q}}{{\overline{\mathit q}}}$ state and can be a candidate for an exotic structure.\\
${{\boldsymbol \psi}{(4390)}}$ formerly known as ${{\mathit X}{(4390)}}$ is the latest observed state at BESIII \cite{BESIII:2016adj} in $2017$ during the process $e^+ e^- \to \pi^+ \pi^- h_c$, at center-of-mass energies from $3.896$ to $4.600$ GeV having mass $4391.5{}^{+6.3}_{-6.8}\pm1.0$ MeV and $J^{PC}=1^{- -}$. This state can also show property different from conventional charmonium state and has sparse theoretical and practical knowledge.\\
${{\boldsymbol \psi}{(4660)}}$ previously known as ${{\mathit Y}{(4660)}}$ discovered by Belle \cite{Wang:2007ea,Wang:2014hta} and confirmed by BaBar \cite{Lees:2012pv} and having negative parity.\\
${{\boldsymbol \psi}{(3770)}}$ resonance is a vector state first detected at SPEAR \cite{Rapidis:1977cv} in 1977, PDG \cite{Tanabashi:2018oca} estimates its mass as $3773.13 \pm0.35$ MeV. Godfrey \cite{Godfrey:1985xj} in $1985$ assigned it as $1^3D_1$ state.\\
${{\boldsymbol \psi}{(4160)}}$ having $J^{PC}=1^{- -}$ first experiment evidence given by \cite{Brandelik:1978ei}, recently observed by Belle \cite{Wang:2012bgc} and LHCB \cite{PhysRevD.87.051101} in $2013$. PDG \cite{Tanabashi:2018oca} estimates mass as $4191\pm5$ MeV.\\
${{\mathit Y}{(4260)}}$ and ${{\mathit Y}{(4360)}}$ which have been renamed in PDG \cite{Tanabashi:2018oca} as ${{\boldsymbol \psi}{(4260)}}$  and ${{\boldsymbol \psi}{(4360)}}$ were first observed at BABAR \cite{Aubert:2005rm} in $2005$ and at Belle \cite{Wang:2007ea} in $2007$ respectively. Both are vector states, yet unlike most conventional charmonium do not corresponds to enhancements in ${{\mathit e}^{+}}{{\mathit e}^{-}}$ hadronic cross section nor decay to $D\bar{D}$ but decay as $\pi^+\pi^- \boldsymbol J / \psi$ and $\pi^+\pi^-\boldsymbol\psi(2S)$ respectively. Both these states can have properties different from conventional charmonia state, \cite{Cleven:2013mka} has considered ${{\mathit Y}{(4260)}}$ as a molecular structure. Recently these two states have been studied as S-D states admixtures in \cite{Bhavsar:2018umj} as study of charmonium in relativistic Dirac formalism with linear confinement potential. The above mentioned states have been tabulated in Table\ref{Table:intro}.\\
Charmomium is most dense system in the entire heavy flavor spectroscopy having $37$ experimentally discovered states. In phenomenology the charmonium mass spectrum is computed by many potential models like relativistic quark model \cite{Lakhina:2006vg}, screened potential model \cite{Deng:2016stx,Deng:2016ktl}, constituent quark model \cite{Segovia:2016xqb} and some non-linear potential models  \cite{Patel:2015ywa,Bonati:2015dka,Gutsche:2014oua,Shah:2012js,Negash:2015rua,Li:2009zu,Li:2009nr,Quigg:1977dd,Martin:1980jx,Buchmuller:1980su}.
Cornell potential is most commonly studied potential for heavy quarkonia system and has been supported by lattice QCD simulations as well \cite{Bali:1994de,Bali:2000vr}. Detailed explanation about quark model hypothesis has been discussed in \cite{Raghav:2018bqo}. Quark model for studying heavy quarkonia has some common  features when compared with QCD but it is not a complete QCD approach, hence most forms of QCD inspired potential would result in uncertainties in the computation of the spectroscopic properties particularly in the intermediate range. Different potential models may produce similar mass spectra matching with the experimentally determined masses but the decay properties mainly leptonic decays or radiative transitions are not in agreement with experimental values. Therefore, test for any model is to reproduce mass spectra along with decay properties. Here, we couple Cornell potential with non-relativistic effective field theory(pNRQCD) \cite{Brambilla:2004jw}. The small velocity of charm quark in $c\bar c$ bound state enables us to use non-relativistic effective field theory within QCD to study charmonia. There are three well defined scales in heavy quarkonia namely; hard scale, soft scale and ultarsoft scale and they also follow a well defined pecking order  $m_Q\gg mv \gg mv^2$, with $m \gg \Lambda_{QCD}$, $\Lambda_{QCD}$ is a QCD scale parameter. NRQCD cannot distinguish soft and ultrasoft scales, which complicates power counting. pNRQCD \cite{Pineda:1997bj,Brambilla:1999xf} solves this problem by integrating the energy scales above $mv$ in NRQCD. The above statements have been discussed in detail in previous work \cite{Raghav:2018bqo}. Recently a study on Cornell Model calibration with NRQCD at $N^3LO$ has been done \cite{Ortega:2018oae}. A spin dependent relativistic correction term (in the framework of pNRQCD \cite{koma}) has been added with Coulomb plus confinement potential in the present work, and the Schr\"{o}dinger equation has been solved numerically \cite{Lucha:1998xc}.\\
The theoretical framework has been discussed in Section \ref{sec:th}, Various decays of S and P wave states has been discussed in Section \ref{sec:decay}, Charmoniumlike negative and positive parity states have been discussed in Section \ref{sec:mix}, Electromagnetic transition widths are in Section \ref{sec:em} and finally results, discussion and conclusion are presented in Sections \ref{sec:result} and \ref{sec:conclusion}.
\begin{table*}[!htb]
\caption{Experimental status of some negative and positive parity $c\bar c$ mesons near open-flavor threshold reported by PDG\cite{Tanabashi:2018oca}}.
%\begin{center}
\begin{tabular*}{\textwidth}{@{\extracolsep{\fill}}cccccc}
\hline \hline
PDG& Former/Common  & Expt.Mass  & J$^P$ & Production & Discovery\\
Name&Name&(in keV)&&&Year\\
\hline
${{\boldsymbol \psi}{(3770)}}$&--&3773.13$\pm$0.35&$1^-$&${{\mathit e}^{+}}{{\mathit e}^{-}}\rightarrow{{\mathit D}}{{\overline{\mathit D}}}$&2012\\

${{\boldsymbol \chi}_{{c1}}{(3872)}}$&${{\mathit X}{(3872)}}$&3871.69$\pm$0.17&$1^+$&${{X\mathit B}}\rightarrow{{\mathit K}}{ \pi}^{+}{\pi}^{-}{{\mathit J / \psi}{(1S)}}$&2003\\

${{\mathit X}{(3900)}}$&${{\mathit X}{(3900)}}$&3886.6$\pm$2.4&$1^+$&${{\boldsymbol \psi}{(4260)}}\rightarrow{\pi}^{-}{{\mathit X}}$&2013\\

${{\mathit X}{(3915)}}$ & ${{\boldsymbol \chi}_{{c0}}{(3915)}}$ & 3918.4$\pm$1.9 & $0^{+}or2^{+}$ & ${{\mathit e}^{+}}{{\mathit e}^{-}}\rightarrow{{\mathit e}^{+}}{{\mathit e}^{-}}{{\mathit X}}$&2004\\

${{\boldsymbol \chi}_{{c1}}{(4140)}}$&${{\mathit X}{(4140)}}$&4146.8$\pm$2.4&$1^{+ }$&${{\mathit B}^{+}}\rightarrow{{\boldsymbol \chi}_{{c1}}}{{\mathit K}^{+}}$&2008\\
&&&&${{\mathit e}^{+}}{{\mathit e}^{-}}\rightarrow{{\mathit e}^{+}}{{\mathit e}^{-}}{{\mathit X}}$&\\

${{\boldsymbol \psi}{(4160)}}$&--&4191$\pm$5&$1^{- }$&${{\mathit e}^{+}}{{\mathit e}^{-}}\rightarrow{{\mathit J / \psi}}{{\mathit X}}$&2007\\

${{\boldsymbol \psi}{(4230)}}$&${{\mathit X}{(4230)}}$&$4218 {}^{+5}_{-4}$&$1^{-}$&${{\mathit e}^{+}}{{\mathit e}^{-}}\rightarrow{{\mathit X}}$&2015\\

${{\boldsymbol \psi}{(4260)}}$&${{\mathit Y}{(4260)}}$&4230$\pm$8&$1^{- }$&${{\mathit e}^{+}}{{\mathit e}^{-}}\rightarrow{{\mathit X}}$&2005\\
&${{\mathit X}{(4260)}}$&&&&\\

${{\boldsymbol \chi}_{{c1}}{(4274)}}$&${{\mathit X}{(4274)}}$&$4274{}^{+8}_{-6}$&$1^{+ }$&${{\mathit{B}}^{+}}\rightarrow{{\mathit {K}}^{+}}{{\mathit X}}$&2011\\

${{\boldsymbol \psi}{(4360)}}$&${{\mathit Y}{(4360)}}$&4368$\pm$13&$1^{- }$&${{\mathit e}^{+}}{{\mathit e}^{-}}\rightarrow{{\mathit X}}$&2007\\

${{\boldsymbol \psi}{(4390)}}$&${{\mathit X}{(4390)}}$&4392$\pm$7&$1^{- }$&${{\mathit e}^{+}}{{\mathit e}^{-}}\rightarrow{{\mathit X}}$&2017\\

${{\boldsymbol \psi}{(4660)}}$&${{\mathit Y}{(4660)}}$&$4643 \pm9$&$1^{- }$&${{\mathit e}^{+}}{{\mathit e}^{-}}\rightarrow{{\mathit X}}$&2007\\
\hline \hline
\end{tabular*}%\end{center}
$\label{Table:intro}$
\end{table*}

\section{Theoretical framework}
\label{sec:th}
Considering charmonium as non-relativistic system we use the following Hamiltonian to calculate its mass spectra. Same theoretical framework has been used by us for studding bottomonium in effective filed theory formalism\cite{Raghav:2018bqo}.
\begin{equation}
  H=M+\frac{P^2}{2 \mu}+ V_{pNRQCD}(r)+ V_{SD}(r)
  \end{equation}
Here, $M$ and $\mu$ represents the total and the reduced mass of the system.
Three terms namely Coulombic term $V_v$(r) (vector), a confinement term $V_s$ (scalar) and relativistic correction $V_p(r)$ in the framework of pNRQCD\cite{Brambilla:2000gk,article,Raghav:2018bqo,PerezNadal:2008vm} has been included in the interaction potential $V_{pNRQCD}(r)$. After fitting the spin average ground state mass(1S) with its experimental value, we fix the mass of charm quark and other potential parameters like $\epsilon$, $A$, $\alpha_s$, $\sigma$, $C$ and $a$, there values are given in Table \ref{Table:parameters}, With these parameters, $\chi^2$/\textit{d.o.f}\cite{Huang:2019agb} is estimated
to be $1.553$. Using these fixed values we generate the entire mass spectrum of charmonium by solving the Schr\"{o}dinger equation numerically \cite{Lucha:1998xc}.
\begin{eqnarray}
V_{pNRQCD}(r) & = & V_v(r) + V_s(r) + V_p(r) \cr \nonumber\\
 V_{pNRQCD}(r) &= & -\frac{4\alpha_{s}}{3r}+ Ar + \frac{1}{m_c}V^{(1)}(r) + \frac{1}{m_c^2}V^{(2)}(r)
\end{eqnarray}
\begin{eqnarray}
V^{(1)}(r)= -\frac{9\alpha_{c}^2}{8r^2}+ a \log r + C
\end{eqnarray}
The parameter $A$ represents potential strength analogous to spring tension. $\alpha_s$ and $\alpha_c$ are strong and effective running coupling constants respectively, $m_{c}$ is mass of charm quark, $a$ and $C$ are potential parameters. Spin-dependent part of the usual one gluon exchange potential has been considered to obtain mass difference between degenerate mesonic states,

\begin{eqnarray}
V^{(2)}(r) & = & V_{SS}(r)\bigg[S(S+1)-\frac{3}{2}\bigg]+ V_{L\cdot S}(r)(\overrightarrow{L}\cdot \overrightarrow{S}) \cr && +
V_T(r) \bigg [ S(S+1)-3(S\cdot \hat{r})(S\cdot \hat{r}) \bigg]
\end{eqnarray}
Where the spin-spin interaction
\begin{eqnarray}
V_{SS}(r) = \frac{8}{9 }\frac{\alpha_s}{m_c^2}\overrightarrow{S}_Q \overrightarrow{S}_{\bar{Q}} 4 \pi \delta^3(\vec r),
\end{eqnarray}
and the spin-orbital interaction
\begin{eqnarray}
V_{L.S}(r) &=& \frac{1}{m_c^2} \bigg( \frac{C_s}{2 r} \frac{d}{dr} (V_v(r) + V_s(r)) \cr && + \frac{C_f}{r}\left[-\left(1-\epsilon\right)A +\left(\frac{\alpha_c}{r^2}+ \epsilon A \right)\right] \bigg)
\end{eqnarray}
with $C_f=\frac{4}{3}$, and the tensor interaction
\begin{eqnarray}
V_{T}(r) &=& \frac{1}{m_c^2}\frac{{C_f}^2}{3} \frac{3 \alpha }{r^3}
%S_{12} &=& 3 \frac{(s_1 . r_1)(s_2 . r_2)}{r^2} - s_1.s_2
\end{eqnarray}

\begin{table*}[!htb]
\caption{Potential parameters}
%\renewcommand{\arraystretch}{0.7}
%\begin{center}
\begin{tabular*}{\textwidth}{@{\extracolsep{\fill}}ccccccc}
\hline\hline
$\alpha_c$ & $m_c$ & $\epsilon$ & A & $\alpha_s$ &  C & a \\
\hline
0.4& 1.321GeV & 0.12 & 0.191 $\frac{GeV}{fm}$ & 0.318 & 0.12 &  -0.165 ${GeV}^2$ \\
\hline
&&&$\chi^2$/\textit{d.o.f}=1.553\\
\hline\hline
\end{tabular*}%\end{center}
\label{Table:parameters}
\end{table*}
Here, the effect of relativistic corrections, the ${\cal O}\left(1/m\right)$ correction, the spin-spin, spin-orbit and tensor corrections ${\cal O}\left(1/m^2\right)$ are tested for charmonium.\\
The computed masses of $S$, $P$, $D$ and $F$ states are tabulated in Tables \ref{Table:mass1} and \ref{Table:mass2}, along with latest experimental data and other theoretical approaches and is found to be in good agreement with them.

\begin{table*}[!htb]
\caption{S and P state mass spectra of $c\bar c$ meson (in GeV)}
%\begin{center}
\begin{tabular*}{\textwidth}{@{\extracolsep{\fill}}cccccccc}
\hline \hline
State &Present& \cite{Tanabashi:2018oca}&\cite{Chaturvedi:2018xrg}& \cite{Ebert:2011jc}& \cite{Sultan:2014oua} & \cite{Soni:2017wvy}& LQCD\cite{Kalinowski:2015bwa} \\
\hline
$1^1S_0$ &2.989&2.984$\pm$0.005 &3.004& 2.981  &2.982&2.989&2.884\\
$1^3S_1$ &3.094&3.097$\pm$0.006 &3.086& 3.096  &3.090&3.094&3.056\\
\hline

$2^1S_0$ &3.572&3.639$\pm$0.012 &3.645& 3.635  &3.630&3.602&3.535\\
$2^3S_1$ &3.649&3.686$\pm$0.025 &3.708& 3.685  &3.672&3.681&3.662\\
\hline

$3^1S_0$ &3.998& --    & 3.989 & 4.043 &4.058 &--&--\\
$3^3S_1$ &4.062& 4.039$\pm$0.043& 4.039 & 4.072&4.129&--&--\\
\hline

$4^1S_0$ &4.372& --    &4.534& 4.401  &4.384&4.448&-- \\
$4^3S_1$ &4.428& 4.421$\pm$0.004&4.579& 4.427 &4.406&4.514&--  \\
\hline

$5^1S_0$ &4.714& --    &4.901& 4.811 &4.685&4.799&--  \\
$5^3S_1$ &4.763& --  &4.942& 4.837 & 4.704&4.863&-- \\
\hline
$6^1S_0$ &5.033& --    &5.240&5.151  &4.960&5.124&--   \\
$6^3S_1$ &5.075& --  &5.277& 5.167   & 4.977&5.185&--\\

\hline
$1^3P_0$ &3.473& 3.414$\pm$0.031 &3.440& 3.413  &3.424&3.428&3.421\\
$1^3P_1$ &3.506& 3.510$\pm$0.007 &3.492& 3.511  &3.505&3.468&3.480\\
$1^1P_1$ &3.527& 3.525$\pm$0.038 &3.496& 3.525  &3.516&3.470&3.494\\
$1^3P_2$ &3.551& 3.556$\pm$0.007 &3.511& 3.555& 3.549&3.480&3.536\\
\hline

$2^3P_0$ &3.918& 3.918 $\pm$0.019$^\star$ &3.932& 3.870 & 3.852&3.897&--\\
$2^3P_1$ &3.949& 3.871 $\pm$0.001$^\star$  &3.984& 3.906 &3.925&3.938&--\\
$2^1P_1$ &3.975& --    &3.991& 3.926 &3.934&3.943&--\\
$2^3P_2$ &4.002& 3.927$\pm$.026 &4.007& 3.949 &3.965&3.955&--\\
\hline

$3^3P_0$ &4.306&--&4.394& 4.301 &4.202&4.296&--\\
$3^3P_1$ &4.336&--&4.401& 4.319 &4.271&4.338&--\\
$3^1P_1$ &4.364&--&4.410& 4.337 &4.279&4.344&--\\
$3^3P_2$ &4.392&--&4.427& 4.354  &4.309 &4.358&--\\
\hline

$4^3P_0$ &4.659& --&4.722& 4.698 &4.509&4.653&--\\
$4^3P_1$ &4.688& --&4.771& 4.728 &4.576&4.696&--\\
$4^1P_1$ &4.716& --&4.784& 4.744 &4.585&4.704&--\\
$4^3P_2$ &4.744& --&4.802& 4.763 &4.614&4.718&--\\
\hline \hline
\end{tabular*}%\end{center}
\label{Table:mass1}
\end{table*}

\begin{table*}[!htb]
\caption{D and F wave mass spectra of $c\bar c$ meson (in GeV)}
%\begin{center}
\begin{tabular*}{\textwidth}{@{\extracolsep{\fill}}ccccccc}
\hline \hline
State&& \cite{Tanabashi:2018oca}&\cite{Chaturvedi:2018xrg}&\cite{Ebert:2011jc}&\cite{Sultan:2014oua}&\cite{Soni:2017wvy}\\
\hline
$1^3D_3$ &3.806& --&3.798& 3.813 & 3.805&3.755\\
$1^3D_2$ &3.800& 3.822$\pm$0.012 \cite{Bhardwaj:2013rmw} &3.814&3.795 &3.800&3.772\\
$1^3D_1$ &3.785& 3.773$\pm$0.035$^\star$&3.815& 3.783  &3.785&3.775\\
$1^1D_2$ &3.780& --&3.806& 3.807&3.799&3.765\\
\hline

$2^3D_3$ &4.206& --&4.273& 4.220 &4.165&4.176\\
$2^3D_2$ &4.203& --&4.248& 4.190 &4.158&4.188\\
$2^3D_1$ &4.196& 4.191$\pm$0.005$^\star$ &4.245& 4.105 &4.141&4.188\\
$2^1D_2$ &4.203& --&4.242& 4.196  &4.158&4.182\\
\hline

$3^3D_3$ &4.568& --&4.626& 4.574 &4.481&4.549\\
$3^3D_2$ &4.566& --&4.632& 4.544 &4.472&4.557\\
$3^3D_1$ &4.562& --&4.627& 4.507 &4.455&4.555\\
$3^1D_2$ &4.566& --&4.629& 4.549 &4.472&4.553\\
\hline

$4^3D_3$ &4.902&--&4.920&--&--&4.890\\
$4^3D_2$ &4.901&--&4.896&--&--&4.896\\
$4^3D_1$ &4.898&--&4.857&--&--&4.891\\
$4^1D_2$ &4.901&--&4.898&--&--&4.892\\
\hline

%$5^3D_3$ &5.216& --    &&  &&&--\\
%$5^3D_2$ &5.215& --    &&  &&&--\\
%$5^3D_1$ &5.213& --    &&  &&&--\\
%$5^1D_2$ &5.215& --    &&  &&&--\\
%\hline
$1^3F_2$&4.015&--&4.041&--&--&3.990\\
$1^3F_3$&4.039&--&4.068&--&--&4.012\\
$1^3F_4$&4.052&--&4.093&--&--&4.036\\
$1^1F_3$&4.039&--&4.071&--&--&4.017\\
\hline
$2^3F_2$&4.403&--&4.361&--&--&4.378\\
$2^3F_3$&4.413&--&4.400&--&--&4.396\\
$2^3F_4$&4.418&--&4.434&--&--&4.415\\
$2^1F_3$&4.413&--&4.406&--&--&4.400\\
\hline
$3^3F_2$&4.751&--&--&--&--&4.730\\
$3^3F_3$&4.756&--&--&--&--&4.746\\
$3^3F_4$&4.759&--&--&--&--&4.761\\
$3^1F_3$&4.756&--&--&--&--&4.749\\
\hline \hline
\end{tabular*}%\end{center}
\label{Table:mass2}
\end{table*}

\section{Decay Widths of $S$ and $P$ charmoium states using NRQCD approach}
\label{sec:decay}
Successful determination of decay widths along with the mass spectrum calculation is very important for  believability of any potential model. Better insight into quark gluon dynamics can be provided by studding strong decays, radiative decays and leptonic decays of vector mesons. Determined radial wave functions and extracted model parameters are utilized to compute various decay widths. The short distance and long distance factors in NRQCD are calculated in terms of running coupling constant and non-relativistic wavefunction.

\subsection{$\gamma\gamma$ decay width}
The $\gamma\gamma$ decay widths of $S$-wave states have been calculated at NNLO in $\nu$, at NLO in $\nu^2$, at $O(\alpha_s \nu^2)$ and at NLO in $\nu^4$.
NRQCD factorization expression for the decay widths of quarkonia at NLO in $\nu^4$ is given as\cite{Braaten:1995ej}
\begin{eqnarray}\label{eq:nq1}
\nonumber && \Gamma(^1 S_0 \rightarrow \gamma \gamma)  =
\frac{F_{\gamma
\gamma}(^1S_0)}{m^2_Q} \left |\left<0|\chi^{\dag}\psi|^1 S_0\right>\right|^2 \\
 \nonumber  && +
\frac{G_{\gamma \gamma}(^1S_0)}{m^4_Q} Re \left
[\left<^1S_0|\psi^{\dag}\chi|0\right>\left<0|\chi^{\dag}\left(-\frac{i}{2}\overrightarrow{D}\right)^2\psi|^1S_0\right>\right]
\cr \\
\nonumber   &&+ \frac{H^1_{\gamma \gamma}(^1S_0)}{m^6_Q}\left<^1
S_0|\psi^{\dag}\left(-\frac{i}{2}\overrightarrow{D}\right)^2\chi|0\right> \times\\
&& \left<0|\chi^{\dag}\left(-\frac{i}{2}\overrightarrow{D}\right)^2\psi|^1S_0\right> + \frac{H^2_{\gamma \gamma}(^1S_0)}{m^6_Q}\times \nonumber \\ &&
   Re\left[\left<^1S_0|\psi^{\dag}\chi|0\right>\left<0|\chi^{\dag}\left(-\frac{i}{2}\overrightarrow{D}\right)^4\psi|^1S_0\right>\right]
\end{eqnarray}
The matrix elements that contribute to the decay rates of the S wave states to $\gamma \gamma$ are given as,
\begin{eqnarray}\label{eq:me1}
&&\left<^1S_0|{\cal{O}}(^1S_0)|^1S_0\right>=\left|\left<0|\chi^{\dag}\psi|^1S_0\right>\right|^2[1+
O(v^4 \Gamma)]\nonumber \\
&&\left<^3S_1|{\cal{O}}(^3S_1)|^3S_1\right>=\left|\left<0|\chi^{\dag}\sigma\psi|^3S_1\right>\right|^2[1+O(v^4 \Gamma)]\nonumber \\
&&\left<^1S_0|{\cal{P}}_1(^1S_0)|^1S_0\right>=Re\left[\left<^1S_0|\psi^{\dag}\chi|0\right> \right. \times \nonumber \\
&&\left. \left<0|\chi^{\dag}(-\frac{i}{2}\overrightarrow{D})^2 \psi|^1S_0\right>\right]+ O(v^4\Gamma )
\end{eqnarray}
The matrix elements are expressed in terms of the regularized wave-function parameters\cite{Bodwin:1994jh}
\begin{eqnarray}\label{eq:wf1}
 &&\left<^1S_0|{\cal{O}}(^1S_0)|^1S_0\right>=\frac{3}{2 \pi}|R_{P}(0)|^2 \nonumber \\
&&\left<^1S_0|{\cal{P}}_1(^1 S_0)|^1S_0\right>
=-\frac{3}{2 \pi}|\overline{R^*_{P}}\ \overline{\bigtriangledown^2 R_{P}}| \nonumber \\
&&\left<^1S_0|{\cal{Q}}^1_1(^1S_0)|^1S_0\right>=
-\sqrt{\frac{3}{2\pi}} \overline{\nabla^2} R_{P}\nonumber \\
\end{eqnarray}
From equation \ref{eq:nq1}, for calculations at leading orders in $\nu$ only the first term is considered, for calculation at leading orders at $\nu^2$ the first two terms are considered, and for calculation at leading orders at $\nu^4$ all the terms are considered.
The coefficients $F,G \& H$ are written as\cite{Brambilla:2010cs,Bodwin:2002hg,Feng:2015uha,Brambilla:2018tyu,Bodwin:1994jh,Jia:2011ah}
\begin{eqnarray}\label{eq:nq2}
&&F_{\gamma \gamma}(^1 S_0)=2\pi Q^4 \alpha^2 \bigg[ 1 + C_F {\alpha_s\over \pi}\left( \frac{\pi^2}{4}-5 \right)\nonumber \\
&& + C_F{\alpha_s^2\over \pi^2}\bigg[C_F \left(-21-\pi^2 \left(\frac{1}{4 \epsilon} + ln\frac{\mu}{m} \right) \right) + \nonumber \\ &&
C_A \left(-4.79 - \frac{\pi^2}{2}\left( \frac{1}{4 \epsilon} + ln\frac{\mu}{m} \right) \right) - \nonumber \\ &&
 0.565N_LT_R + 0.22N_HT_R \bigg] \bigg]
\end{eqnarray}
\begin{eqnarray}\label{eq:nq3}
G_{\gamma \gamma}(^1S_0)=
-{8\pi Q^4\alpha^2 \over 3}\bigg[1+ {C_F\alpha_s \over \pi}\nonumber \\ &&
\bigg({5\pi^2\over 16}-{49\over 12}- \ln {\mu^2 \over 4m^2}\bigg)\bigg]
\end{eqnarray}
\begin{eqnarray}\label{eq:nq4}
H_{\gamma \gamma}(^1S_0)+H^2_{\gamma\gamma}(^1S_0)=\frac{136\pi}{45} Q^4 \alpha^2
\end{eqnarray}
Here, $C_F=4/3$, $C_A=3$, $N_H=1$, $T_R=1/2$, $N_L=3$ and $\mu=0.5$.
For calculations at NNLO in $\nu$, the entire equation \ref{eq:nq2} is used. For calculations at NLO in $\nu^2$ only the first two terms in the square bracket of equation \ref{eq:nq2} is used, and $G_{\gamma \gamma}(^1S_0)$ is taken as $-\frac{8 \pi Q^4}{3}\alpha^2$. For NLO in $\nu^4$, in addition to first two terms in the square bracket of equation \ref{eq:nq2} and $G_{\gamma \gamma}(^1S_0)= -\frac{8 \pi Q^4}{3}\alpha^2$, equation \ref{eq:nq4} is also used. But, for calculation at $O(\alpha_s \nu^2)$  the first two terms in square bracket of equation \ref{eq:nq2} and the entire equation \ref{eq:nq4} are only used. The calculated decay widths are tabulated in table\ref{Table:gammas}.\\

The decay widths for $n^3P_J (J=0,2)$ states to NLO in $\nu^2$ and NNLO in $\nu^2$ have also been calculated. $\gamma\gamma$ decay width of $n^3P_0$ and $n^3P_2$ is expressed as,
\begin{eqnarray}\label{eq:p1}
 \Gamma(\chi_{cJ} \rightarrow \gamma \gamma) &=& \frac{3 N_c Im F_{\gamma\gamma} (^3P_J)}{\pi m_Q^4},\ \ \  J = 0,2.
\end{eqnarray}
Short distance coefficients $F$'s, at NNLO in $\nu^2$ are given by \cite{Sang:2015uxg,Bodwin:2002hg}
%\begin{eqnarray}
%F_{\gamma\gamma} (^3P_0) &=& 3 \pi Q^4 \alpha^2 \left[ 1 + \left( \frac{{\pi}^2}{4} - \frac{7}{3} %\right) C_F \frac{\alpha_s}{\pi} \right] \\
%F_{\gamma\gamma} (^3P_2) &=& \frac{4 \pi Q^4 \alpha^2}{5} \left[1 - 4 C_f \frac{\alpha_s}{\pi} \right]
%\end{eqnarray}
%Short distance coefficients $F$'s, at NLO in $\nu^2$ are given by\cite{Bodwin:2002hg}
\begin{eqnarray}\label{eq:p2}
F_{\gamma\gamma} (^3P_0) &=& 3 \pi Q^4 \alpha^2 \bigg[1+C_F {\alpha_s\over \pi}\bigg(\frac{\pi^2}{4} -\frac{7}{3}\bigg)
\nonumber\\
&&
+ \frac{\alpha_s^2}{\pi^2}\bigg[C_F\frac{\beta_0}{4}\bigg(\frac{\pi^2}{4}-\frac{7}{3}\bigg)
\ln\frac{\mu_R^2}{m^2}\bigg]\bigg]
\end{eqnarray}
\begin{eqnarray}
F_{\gamma\gamma} (^3P_2) &=& \frac{4 \pi Q^4 \alpha^2}{5}
\bigg[1 - 4 C_f \frac{\alpha_s}{\pi} +
 \nonumber\\
&&
\frac{\alpha_s^2}{\pi^2} \bigg(-2C_F\frac{\beta_0}{4}\ln\frac{\mu_R^2}{m^2}\bigg) \bigg]
\end{eqnarray}
$\beta_0 = {11\over 3}C_A - {2\over 3}(n_L+n_H)$ is
the one-loop coefficient of the QCD $\beta$-function, where
$n_H=1$, $C_A=3$ and $n_L$ signifies the number of light quark flavors ($n_L=3$ for $\chi_c$). The calculated decay widths are tabulated in table\ref{Table:gammap}.

\begin{table*}[!htb]
\caption{$\gamma\gamma$ decay widths of $n\eta_c$ meson(in keV)}\
\centering
%\begin{center}
\begin{tabular*}{\textwidth}{@{\extracolsep{\fill}}ccccccc}
\hline \hline
$\Gamma$ &&& State &&& \\
\hline
&$1^1S_0$&$2^1S_0$&$3^1S_0$&$4^1S_0$&$5^1S_0$&$6^1S_0$\\
\hline
NNLO in $\nu$&12.4&8.257&7.102&6.474&6.093&5.788\\
NLO in $\nu^2$&14.129&9.273&6.238&4.674&3.747&3.125\\
$O(\alpha_s \nu^2)$&14.897&9.734&6.547&4.906&3.933&3.280\\
NLO in $\nu^4$&6.725&3.178&1.493&0.858&0.560&0.394\\
\cite{Tanabashi:2018oca}&5.05$\pm0.01$&2.14$\pm$0.04&&&\\
\cite{Chaturvedi:2018xrg}&8.246&4.560&3.737&3.340&3.095&2.924\\
\cite{Soni:2017wvy}&5.618&2.944&2.095&1.644&1.358&1.158\\
\cite{Lakhina:2006vg}&7.18&1.71&1.21&\\
\cite{Ebert:2002pp}&5.5&1.8&\\
\cite{Lansberg:2006dw}&7.5-10&&&&\\
\cite{Kim:2004rz}&7.14$\pm$0.95&4.4$\pm$0.48&&&\\
\hline \hline
\end{tabular*}%\end{center}
\label{Table:gammas}
\end{table*}

\begin{table*}[!htb]
\caption{$\gamma\gamma$ decay widths of $n^3P_J (J=0,2)$ (in keV)}\
%\begin{center}
\begin{tabular*}{\textwidth}{@{\extracolsep{\fill}}ccccccccc}
\hline \hline
$\Gamma$ &&$\chi_0$&&&&$\chi_2$& \\
\hline
$\Gamma$ &1P&2P&3P&4P&1P&2P&3P&4P \\
\hline
NLO in $\nu^2$&4.185&4.306&4.847&4.346&0.538&0.554&0.626&0.559\\
NNLO in $\nu^2$&4.134&4.263&4.799&4.303&0.868&0.893&1.005&0.901\\
\cite{Tanabashi:2018oca}&2.341$\pm$0.189&&&&0.528$\pm$0.404&\\
\cite{Amsler:2008zzb}&2.87$\pm$0.39&&&&0.53$\pm$0.05&\\
\cite{Chaturvedi:2018xrg}&2.692&4.716&8.078&0.928&1.242&1.485&1.691&1.721\\
\cite{Hwang:2010iq}&2.36$\pm$0.35&&&&0.346$\pm$0.009&0.23&\\
\cite{Gupta:1996ak}&6.38&&&&0.57&\\
%\cite{B.Patel}&7.33&8.70&&&1.95&2.32&&\\
\cite{Huang:1996cs}&3.72$\pm$1.1&&&&0.490$\pm$0.150\\
\hline \hline
\end{tabular*}%\end{center}
\label{Table:gammap}
\end{table*}

\subsection{$e^{+}e^{-}$ decay width}
The $e^{+}e^{-}$ decay width of $S$-wave states have been calculated at NLO in $\nu$, NNLO in $\nu$, NLO in $\nu^2$, NLO in $\alpha_s \nu^4$ and NLO in $\alpha_s^2 \nu^4$. NRQCD factorization expression for the decay widths of quarkonia at NNLO in $\nu^4$ is written as,
\begin{eqnarray} \label{eq:nq5}
&&\Gamma(^3S_1 \rightarrow
e^+e^-) = \frac{F_{ee}(^3S_1)}{m^2_Q}
\left|\left<0|\chi^{\dag}\sigma\psi|^3S_1\right>\right|^2 \cr
&&+\frac{G_{ee}(^3S_1)}{m^4_Q}
Re\left[\left<^3S_1|\psi^{\dag}\sigma\chi|0\right>\left<0|\chi^{\dag}\sigma\left(-\frac{i}{2}\overrightarrow{D}\right)^2\psi|^3S_1\right>\right]
\cr
&&+\frac{H^1_{ee}\left(^1S_0\right)}{m^6_Q}
\left<^3S_1|\psi^{\dag}\sigma\left(-\frac{i}{2}\overrightarrow{D}\right)^2\chi|0\right> \times \nonumber \\
&&
\left<0|\chi^{\dag}\sigma\left(-\frac{i}{2}\overrightarrow{D}\right)^2\psi|^3S_1\right>+ \frac{H^2_{ee}(^1 S_0)}{m^6_Q}\times \nonumber \\
&&
\ Re \left[\left<^3S_1|\psi^{\dag}\sigma\chi|0\right>
\left<0|\chi^{\dag}\sigma\left(-\frac{i}{2}\overrightarrow{D}\right)^4\psi|^3S_1\right>\right]
\end{eqnarray}
From equation \ref{eq:nq5}, for calculations at leading orders in $\nu$ only the first term is considered, for calculation at leading orders at $\nu^2$ the first two terms are considered, and for calculation at leading orders at $\nu^4$ all the terms are considered.
The matrix elements that contribute to the decay rates of $\psi \rightarrow e^+e^-$ through the vacuum-saturation approximation gives \cite{Bodwin:1994jh}.
\begin{eqnarray}\label{eq:me2}
&&\left<^3S_1|{\cal{P}}_1(^3S_1)|^3S_1\right>=
Re\left[\left<^3S_1|\psi^{\dag}\sigma\chi|0\right> \right.\times \nonumber\\
&&\left. \left<0|\chi^{\dag}\times \nonumber
\sigma\left(-\frac{i}{2}\overrightarrow{D}\right)^2\psi|^3S_1\right>\right]
+ O(v^4 \Gamma ) \nonumber \\
&&\left<^1S_0|{\cal{Q}}^1_1(^1S_0)|^1S_0\right>=
\left<0|\chi^{\dag}\left(-\frac{i}{2}\overrightarrow{D}\right)^2\psi|^1S_0\right>\nonumber\\
&&\left<^3S_1|{\cal{Q}}^1_1(^3S_1)|^3S_1\right>=\left<0|\chi^{\dag} \sigma
\left(-\frac{i}{2}\overrightarrow{D}\right)^2\psi|^3S_1\right>
\end{eqnarray}
The matrix elements are expressed in terms of the regularized wave-function parameters\cite{Bodwin:1994jh}.
\begin{eqnarray}\label{eq:wf2}
&&\left<^3S_1|{\cal{O}}(^3S_1)|^3S_1\right>=\frac{3}{2\pi}|R_{V}(0)|^2 \nonumber \\
&&\left<^3S_1|{\cal{P}}_1(^3 S_1)|^3S_1\right>=-\frac{3}{2
\pi}|\overline{R^*_{V}}\ \overline{\bigtriangledown^2 R_{V}}| \nonumber \\
&&\left<^3S_1|{\cal{Q}}^1_1(^3S_1)|^3S_1\right>=-
\sqrt{\frac{3}{2\pi}} \overline{ \nabla^2} R_{V} \nonumber \\
\end{eqnarray}
The coefficients $F,G \& H$ are written as\cite{Brambilla:2010cs,Bodwin:2002hg,Marquard:2014pea,Bodwin:2002hg,Bodwin:1994jh}.
\begin{eqnarray}\label{eq:nq6}
&&F_{ee}(^3S_1)= \frac{2 \pi Q^2 \alpha^2}{3}  \bigg[ 1- 4 C_F
\frac{\alpha_s}{\pi}  + \nonumber \\
&&\left[-117.46+0.82n_f+\frac{140\pi^2}{27} ln\left(\frac{2m}{\mu_A}\right)\right]
\left(\frac{\alpha_s}{\pi}\right)^2C_F  \bigg]
\end{eqnarray}
\begin{eqnarray}\label{eq:nq7}
&&G_{ee}(^3 S_1)=-
\frac{8 \pi Q^2}{9} \alpha^2
\end{eqnarray}
\begin{eqnarray}\label{eq:nq8}
H^1_{ee}(^3S_1)+H^2_{ee}(^3S_1)=\frac{58\pi}{54} Q^2 \alpha^2
\end{eqnarray}
For calculations at NLO in $\nu$, only the first two terms from equation \ref{eq:nq6} is used. For calculations at NNLO in $\nu$ the entire equation \ref{eq:nq6} is used. For calculation at $O(\nu^2)$ NLO only the first two terms from equation \ref{eq:nq6} is used and equation \ref{eq:nq7} is also used. For calculation at $O(\alpha_s \nu^4)$ NLO first two terms of equation \ref{eq:nq6}, equations \ref{eq:nq7} and \ref{eq:nq8} are also used. And, for calculation at $O(\alpha_s^2 \nu^4)$ NLO equations \ref{eq:nq6}, \ref{eq:nq7} and \ref{eq:nq8} are used. The calculated decay widths are tabulated in table\ref{Table:ee}.\\
\begin{table*}[!htb]
\caption{$e^{+}e^{-}$ decay widths of $nJ/\psi$(in keV)}\
%\begin{center}
\begin{tabular*}{\textwidth}{@{\extracolsep{\fill}}ccccccc}
\hline \hline
$\Gamma$  &&& State && \\
\hline
&$1^3S_1$&$2^3S_1$&$3^3S_1$&$4^3S_1$&$5^3S_1$&$6^3S_1$\\
\hline
NLO in $\nu$&1.957&1.178&0.969&0.860&0.792&0.741\\
NNLO in $\nu$&0.445&0.267&0.220&0.195&0.180&0.168\\
NLO in $\nu^2$&3.431&2.678&2.394&2.241&2.147&2.075\\
NLO in $\alpha_s\nu^4$&4.100&2.464&2.029&1.801&1.659&1.551\\
NLO in $\alpha_s^2\nu^4$&3.004&3.540&3.350&3.240&3.181&3.138\\
\cite{Tanabashi:2018oca}&5.55$\pm$0.08&2.33$\pm$0.01&0.86$\pm$0.01&&\\
\cite{Yao:2006px}\cite{Amsler:2008zzb}&5.55$\pm$0.14&2.48$\pm$0.06&0.86$\pm$0.07&0.58$\pm$0.07\\
\cite{Chaturvedi:2018xrg}&6.932&3.727&2.994&2.638&2.423&2.275\\
\cite{Soni:2017wvy}&2.925&1.533&1.091&0.856&0.707&0.602\\
\hline \hline
\end{tabular*}%\end{center}
\label{Table:ee}
\end{table*}

\subsection{Light hadron decay width}
The Light hadron decay width through NLO and NNLO in $\nu^2$ is calculated. The methodology for calculation is given as \cite{Braaten:1995ej,Bodwin:1994jh}.
\begin{eqnarray}
\Gamma(^1S_0 \rightarrow LH) = \frac{N_c Imf_1(^1S_0)}{\pi m_Q^2} |\bar{{R_p}}|^2 + \nonumber
\\ \frac{N_c Img_1(^1S_0)}{\pi m_Q^4} Re(\bar{R_p}^* \bar{\nabla^2 R_p})
\end{eqnarray}
The coefficients $F \& G$ for decay width calculation at NNLO at $\nu^2$ are written as\cite{Bodwin:1994jh,Feng:2017hlu},
\begin{eqnarray}\label{eq:nq9}
&&Imf_1(^1S_0)=\frac{\pi C_F \alpha_s^2}{N_c} \bigg[ 1+\frac{\alpha_s}{\pi}\bigg(\frac{\beta_0}{2}ln\frac{\mu_R^2}{4m^2}+\bigg(\frac{\pi^2}{4}-5 \bigg)C_F\nonumber \\
&& + \bigg(\frac{199}{18}-\frac{13\pi^2}{24} \bigg)C_A -\frac{8}{9}n_L - \frac{2n_H}{3}  \bigg)ln2 + \nonumber \\
&&\frac{\alpha_s^2}{\pi^2} \bigg(-50.1 + \frac{3\beta_0^2}{16} ln^2\frac{\mu_R^2}{4m^2} + \nonumber \\
&&\bigg(\frac{\beta_1}{8}+ \frac{3}{4}\beta_0 10.62  \bigg)ln\frac{\mu_R^2}{4m^2}-\nonumber \\
&& \pi^2\bigg(C_F^2 +\frac{C_A C_F}{2}  \bigg)ln\frac{\mu_{\Lambda}^2}{m^2} \bigg) \bigg]
\end{eqnarray}
\begin{eqnarray}\label{eq:nq10}
&&Img_1(^1S_0)=-\frac{4\pi C_F \alpha_s^2}{3N_c}\bigg[ 1+\frac{\alpha_s}{\pi}\bigg(\frac{\beta_0}{2}ln\frac{\mu_R^2}{4m^2}-C_F ln\frac{\mu_{\Lambda}^2}{m^2} \bigg)-\nonumber \\
&&\bigg(\frac{49}{12}-\frac{5\pi^2}{16} -2ln2 \bigg)C_F + \nonumber \\
&& \bigg(\frac{479}{36}-\frac{11\pi^2}{16}\bigg)C_A - \frac{41}{36}n_L- \frac{2n_H}{3}ln2 \bigg]
\end{eqnarray}
Here, $\beta_0=\frac{11}{3}C_A-\frac{4}{3}T_Fn_f$, $T_F=1/2$, $n_f=n_L+n_H$ signifies number of active flavour quark, $n_L=3$, $n_H=1$, $C_F=\frac{N_C^2-1}{2N_C}$, $C_A=N_C=3$, $\mu_R$ is the renormalisation scale,  $\beta_1=\frac{34}{3}C_A^2-\frac{20}{3}C_AT_Fn_f-4C_AT_Fn_f$ is the two-loop coefficient of the QCD $\beta$ function.\\
For decay width calculation at NLO in $\nu^2$ only the first two terms in the square bracket from equation \ref{eq:nq9} is considered and the entire equation \ref{eq:nq10} is considered. The results of the decay width is tabulated in table \ref{Table:lh}
\begin{table*}[!htb]
\caption{Light hadrons(LH) decay width of $n\eta_C$(in MeV)}\
%\begin{center}
\begin{tabular*}{\textwidth}{@{\extracolsep{\fill}}ccccccc}
\hline
$\Gamma$&$1^1S_0$&$2^1S_0$&$3^1S_0$&$4^1S_0$&$5^1S_0$&$6^1S_0$\\
\hline
NLO in $\nu^2$ &14.324&7.926&5.765&4.564&3.791&3.232\\
NNLO in $\nu^2$ &14.473&7.330&5.052&3.892&3.183&2.687\\
\cite{Chaturvedi:2018xrg}&4.407&1.685&1.074&0.791&0.624&0.514\\
\cite{Fabiano:2002bw}&14.38$\pm$1.07$\pm$1.43\\
\hline
\end{tabular*}%\end{center}
\label{Table:lh}
\end{table*}

\subsection{$\gamma\gamma\gamma$ decay width}
$\gamma\gamma\gamma$ decay width of $nJ/\psi$ states given in \cite{Mackenzie:1981sf,Bodwin:1994jh} through NLO in $\nu^2$ is also calculated and is represented by,
\begin{equation}\label{eq:nq11}
\Gamma(^3S_1 \rightarrow \gamma\gamma\gamma) = {8 (\pi^2 - 9) Q^6 \alpha^3 \over 9 \pi m_Q^2}
\Bigg[ 1 \;-\; 9.46(2) C_F {\alpha_s \over \pi} \Bigg]R_v \;
%XX\label{fpsi3gNLO}
\end{equation}
Here, $\alpha = e^2/4 \pi$.
\begin{table*}[!htb]
\caption{$\gamma\gamma\gamma$ decay width of $J/\psi$ and higher $\psi$ states(in eV)}
%\begin{center}
\begin{tabular*}{\textwidth}{@{\extracolsep{\fill}}ccccccc}
\hline \hline
$\Gamma$ &&&& State && \\
\hline
  &$1^1S_0$&$2^1S_0$&$3^1S_0$&$4^1S_0$&$5^1S_0$&$6^1S_0$ \\
\hline
NLO in $\nu^2$&1.022&0.900&0.857&0.832&0.815&0.801\\
\cite{Chaturvedi:2018xrg}&2.997&1.083&1.046&0.487&0.381&0.312\\
\cite{Patrignani:2016xqp} &1.077$\pm$0.006\\
\hline \hline
\end{tabular*}%\end{center}
\label{Table:ggg}
 %\centering \raggedright{Note:- Here, in the \cite{Patrignani:2016xqp} value of $\Gamma_ {\gamma\gamma\gamma}$ is found indirectly by multiplying $\Gamma_{total} = 92.9 \pm 2.8 keV$ and $\frac{\Gamma_i}{\Gamma_{total}}= (1.16\pm 0.22) * 10^{-4}$ }.
\end{table*}

\section{Mixed charmonium states}
\label{sec:mix}
Experimentally, many hadronic states are observed but not all can be identified as pure mesonic states. Some of them have properties different from pure mesonic states and can be identified as admixture of nearby iso-parity states.
The mass of admixture state ($M_{NL}$) is expressed in terms of two states ($nl$ and $n'l'$) discussed recently in\cite{Bhavsar:2018umj} and also in \cite{Badalian:2009bu,Shah:2012js,Radford:2009qi,Shah:2014yma} and references therein.
\begin{equation}
M_{NL} = \mid a^2\mid M_{nl} + ( 1-\mid a^2\mid ) M_{n'l'}
\end{equation}
Where,$\mid a^2\mid$ $=$ $\cos^2 \theta$ and $\theta$ is mixing angle.
${{\boldsymbol \psi}{(4230)}}$, ${{\boldsymbol \psi}{(4260)}}$, ${{\boldsymbol \psi}{(4360)}}$, ${{\boldsymbol \psi}{(4390)}}$ and ${{\boldsymbol \psi}{(4660)}}$
have been studied as S-D admixture states, their calculated masses(in GeV) and leptonic decay width is tabulated in Table \ref{Table:mix1}.
${{\boldsymbol \chi}_{{c1}}{(4274)}}$ and ${{\boldsymbol \chi}_{{c1}}{(4140)}}$  have been studied as admixture of nearby $P$-wave states calculated masses are tabulated in Table \ref{Table:mix2}. The calculated masses and decay width of admixture states is compared with other theoretical and available experimental results \cite{Shah:2014yma,Patel:2015ywa,Yazarloo:2016zer,Bhavsar:2018umj}. \\
Mixed $P$ wave states can be expressed as,
\begin{equation}
|\alpha\rangle=\sqrt{\frac{2}{3}}|^3P_1\rangle+\sqrt{\frac{1}{3}}|^1P_1\rangle
\end{equation}
\begin{equation}
|\beta\rangle=-\sqrt{\frac{1}{3}}|^3P_1\rangle+\sqrt{\frac{2}{3}}|^1P_1\rangle
\end{equation}
Where, $|\alpha\rangle$, $|\beta\rangle$ are states having same parity.
We can write the masses of these states in terms of the predicted masses of pure $P$ wave states ($^3P_1$ and $^1P_1$) as \cite{Shah:2014yma,Patel:2015ywa,Yazarloo:2016zer},

\begin{table*}[!htb]
\caption{Mass spectra and leptonic decay width of S-D wave admixture states(Negative parity)}
%\begin{center}
\begin{tabular*}{\textwidth}{@{\extracolsep{\fill}}cccccccc}
\hline \hline
Expt.& J$^P$ & Mixed & \% mixing  & \multicolumn {2} {c} {Masses mixed state(GeV)}& \multicolumn {2} {c} {$\Gamma_{e^{+}e^{-}}$ mixed state(eV)}\\
\cline{5-8}  state &&state&of S state& {Our} & {Expt.\cite{Tanabashi:2018oca}}&{Our}&{Expt.} {}\\
\hline
${{\boldsymbol \psi}{(4230)}}$& $1^-$ & $3^3S_1$ and $3^3D_1$ &41\% &4.277&$4.218 {}^{+0.005}_{-0.004}$&11.027&2.7$\pm$0.05\cite{Ablikim:2015dlj}\\

${{\boldsymbol \psi}{(4260)}}$& $1^-$ & $3^3S_1$ and $3^3D_1$ &36\%&4.234&4.230$\pm$0.008&6.352&9.2$\pm$1.0\cite{Tanabashi:2018oca}\\

${{\boldsymbol \psi}{(4360)}}$& $1^-$ & $3^3S_1$ and $3^3D_1$ &51\%&4.363&4.368$\pm$0.013&0.651&6.0$\pm$1.0\cite{Lees:2014iua}\\

${{\boldsymbol \psi}{(4390)}}$& $1^-$ & $3^3S_1$ and $3^3D_1$ &54\%&4.389&4.329$\pm$0.007&2.110&--\\

${{\boldsymbol \psi}{(4660)}}$& $1^-$ & $4^3S_1$ and $4^3D_1$ &43\%&4.648&4.643$\pm$0.009&13.892&8.1$\pm$1.1$\pm$1.0\cite{Wang:2014hta}\\

\hline \hline
\end{tabular*}%\end{center}
\label{Table:mix1}
\end{table*}

\begin{table*}[!htb]
\caption{Mass spectra of P wave admixture states(Positive parity)}
%\begin{center}
\begin{tabular*}{\textwidth}{@{\extracolsep{\fill}}ccccc}
\hline \hline
Expt. State &J$^P$& Mixed State Configuration & Our(GeV) & Expt.\cite{Tanabashi:2018oca}\\
\hline
%${{{\mathit Z}_{{c}}{(3900)}}}$ &$1^+$& $2^1P_1$ and $1^3P_1$ & 3.818 & $3.886 \pm0.024$\\
${{\boldsymbol \chi}_{{c1}}{(4140)}}$ &$1^+$& $3^3P_1$ and $2^1P_1$ & 4.094 & $4.146 \pm0.024$\\
%${{{\mathit Z}_{{c}}{(4200)}}}$ &$1^+$& $2^1P_1$ and $3^3P_1$ & 4.215 & $4.196 {}^{+0.035}_{-0.032}$\\
${{\boldsymbol \chi}_{{c1}}{(4274)}}$ &$1^+$& $3^3P_1$ and $3^1P_1$ & 4.301 & $4.274 {}^{+0.008}_{-0.006}$\\
%${{{\mathit Z}_{{c}}{(4430)}}}$ &$1^+$& $3^1P_1$ and $4^3P_1$ & 4.472 & $4.478 {}^{+0.015}_{-0.018}$\\

\hline \hline
\end{tabular*}%\end{center}
\label{Table:mix2}
\end{table*}

\section{Electromagnetic transition widths}
\label{sec:em}
Electromagnetic transitions have been calculated in this article in the framework of pNRQCD and this study can help to understand the non-perturbative expect of QCD.
For ($E1$) transition the selection rules are $\Delta L = \pm 1$ and $\Delta S = 0$ while for ($M1$) transition it is $\Delta L = 0$ and $\Delta S = \pm 1$.
The obtained normalised reduced wave function and parameters used in current work are employed to electromagnetic transition width calculation.
In nonrelativistic limit, the radiative $E1$ and $M1$ transition widths are given by \cite{Brambilla:2010cs,Radford:2009qi,Eichten:1974af,Eichten:1978tg,Soni:2017wvy}
\begin{eqnarray}
 \Gamma(n^{2S+1}L_{iJ_i} \to n^{2S+1}L_{fJ_f} + \gamma) &=& \frac{4 \alpha_e \langle e_Q\rangle ^2\omega^3}{3} (2 J_f + 1) \cr && \times S_{if}^{E1} |M_{if}^{E1}|^2
 \end{eqnarray}
\begin{eqnarray}
\Gamma(n^3S_1 \to {n'}^{1}S_0+ \gamma) = \frac{\alpha_e \mu^2 \omega^3}{3} (2 J_f + 1) S_{if}^{M1} |M_{if}^{M1}|^2
\end{eqnarray}
where, mean charge content $\langle e_Q \rangle$ of the $Q\bar{Q}$ system, magnetic dipole moment $\mu$ and photon energy $\omega$ are given by
\begin{equation}
\langle e_Q \rangle = \left |\frac{m_{\bar{Q}} e_Q - e_{\bar{Q}} m_Q}{m_Q + m_{\bar{Q}}}\right |%|\frac{m_{\bar{Q}} e_Q - m_Q e_{\bar{Q}}}{m_Q + m_{\bar{Q}}|,
\end{equation}
\begin{equation}
\mu = \frac{e_Q}{m_Q} - \frac{e_{\bar{Q}}}{m_{\bar{Q}}}
\end{equation}
and
\begin{equation}
\omega = \frac{M_i^2 - M_f^2}{2 M_i}
\end{equation}
respectively. Also, the symmetric statistical factors are given by
\begin{equation}
S_{if}^{E1} = {\rm max}(L_i, L_f)
\left\{ \begin{array}{ccc} J_i & 1 & J_f \\ L_f & S & L_i \end{array} \right\}^2\\
\end{equation}
and
\begin{equation}
S_{if}^{M1} = 6 (2 S_i + 1) (2 S_f + 1)
\left\{ \begin{array}{ccc} J_i & 1 & J_f \\ S_f & \ell & S_i \end{array} \right\}^2 \left\{ \begin{array}{ccc} 1 & \frac{1}{2} & \frac{1}{2} \\ \frac{1}{2} & S_f & S_i \end{array} \right\}^2.
\end{equation}
The matrix element $|M_{if}|$ for $E1$ and $M1$ transitions can be written as
\begin{equation}
\left |M_{if}^{E1}\right | = \frac{3}{\omega} \left\langle f \left | \frac{\omega r}{2} j_0 \left(\frac{\omega r}{2}\right) - j_1 \left(\frac{\omega r}{2}\right) \right | i \right\rangle
\end{equation}
and
\begin{equation}
\left |M_{if}^{M1}\right | = \left\langle f\left | j_0 \left(\frac{\omega r}{2}\right) \right | i \right\rangle
\end{equation}
The electromagnetic transition widths are listed in tables \ref{Table:em1} \& \ref{Table:em2} and are also compared with experimental results as well as with other theoretical predictions.

\begin{table*}[!htb]
 \caption{Electric dipole (E1) transitions widths of $c\overline{c}$ mesons.(LP = Linear potential model, SP = Screened potential model, NR = Non-relativistic and  RE = Relativistic)($\Gamma$ in KeV) }
%\begin{center}

\begin{tabular*}{\textwidth}{@{\extracolsep{\fill}}cccccccc}
 \hline\hline
 Transition & Present & \cite{Tanabashi:2018oca} & \cite{Kher:2018wtv} & \cite{Cao:2012du}& \cite{Deng:2016stx}& \cite{Sultan:2014oua}&\cite{Soni:2017wvy}\\
 & work & Expt. &  &  & LP(SP)& RE(NR)&\\
 \hline
 %\addlinespace[2pt]
$1^{3}P_{2} \rightarrow 1^{3}S_{1}$ &457.39 &$406\pm31$  & 233.85 & 405 & 327(338)& 437.5(424.5)&157.22\\
$1^{3}P_{1} \rightarrow 1^{3}S_{1}$ &378.33 & $320\pm25$ &189.86 &341 &269 (278) & 329.5(319.5)&146.32\\
$1^{1}P_{1} \rightarrow 1^{1}S_{0}$  &505.69 & -- & 357.83  &473 &361 (373)& 570.5(490.3)&247.97\\
$1^{3}P_{0} \rightarrow 1^{3}S_{1}$ &145.33 &$131\pm14$ & 118.29  & 104 & 141(146)& 159.2(154.5)&112.03\\
 \hline
  %\addlinespace[2pt]
$2^{3}S_{1} \rightarrow 1^{3}P_{2}$ &28.51 &$26\pm1.5$ & 7.07 & 39 & 36(44)& 35.5 (37.9)&62.31 \\
 $2^{3}S_{1} \rightarrow 1^{3}P_{1}$ &28.79 &$27.9\pm1.5$ & 10.39 & 38 & 45(48)&50.9 (54.2)&43.29 \\
 $2^{3}S_{1} \rightarrow 1^{1}P_{1}$  &28.79 & -- & 7.94 & & & &\\
 $2^{3}S_{1} \rightarrow 1^{3}P_{0}$ &31 & $29.8\pm1.5$ &11.93 &29 & 27(26)& 58.8 (62.6)&21.86\\
 $2^{1}S_{0} \rightarrow 1^{3}P_{1}$  &8.31 &  & 9.20 & & & & \\
 $2^{1}S_{0} \rightarrow 1^{1}P_{1}$  &2.69 &   & 6.05 & 56 &49 (52)& 45.2 (49.9)&36.20\\
   \hline
    %\addlinespace[2pt]
 $1^{3}D_{3} \rightarrow 1^{3}P_{2}$  &348.99 &  & 237.51 & 302 & &397.7(271.1)&175.21 \\
 $1^{3}D_{2} \rightarrow 1^{3}P_{2}$  &66.92 &  & 62.34 & 82& 79(82)&96.52(64.06)&50.31 \\
 $1^{3}D_{2} \rightarrow 1^{3}P_{1}$ &103.70 & & 89.18  &301& 281(291)& 438.2(311.2)&165.17 \\
 $1^{3}D_{1} \rightarrow 1^{3}P_{2}$ &13.13 & $<21$ &6.45 & 8.1&5.4 (5.7)& 4.73(4.86)&5.72 \\
 $1^{3}D_{1} \rightarrow 1^{3}P_{1}$ &90.47 & $70\pm17$ &139.52 &153& 115 (111)& 122.8(126.2)&93.77 \\
 $1^{3}D_{1} \rightarrow 1^{3}P_{0}$ &120.66 & $172\pm30$ &343.87 & 362&243 (232)& 394.6(405.4)&161.50\\
 \hline
    %\addlinespace[2pt]
 $2^{3}P_{2} \rightarrow 2^{3}S_{1}$  &346.02 & &281.93   &264 & & 377.1(287.5)&116.32\\
 $2^{3}P_{1} \rightarrow 2^{3}S_{1}$ &219.71 &  & 206.87 & 234& & 246.0(185.3)&102.67\\
 $2^{1}P_{1} \rightarrow 2^{1}S_{0}$  &493.45 &  & 343.55 &274& & 349.8(272.9)&163.64 \\
 $2^{3}P_{0} \rightarrow 2^{3}S_{1}$  &161.07 &  & 102.23 &83& & 108.3(65.3)&70.40\\
 \hline
  %\addlinespace[2pt]
 $2^{3}P_{2} \rightarrow 1^{3}D_{3}$  &6.88 &  & 33.27 &  76& & 60.67(78.69)&\\
 $2^{3}P_{2} \rightarrow 1^{3}D_{2}$  &7.47 &  & 5.49 & 10& & 11.48(15.34)&\\
 $2^{3}P_{2} \rightarrow 1^{1}D_{2}$  &9.63 &  & 5.83 &&& & \\
 $2^{3}P_{2} \rightarrow 1^{3}D_{1}$ &9.07 &  & 0.41 &0.64& & 2.31(1.67)&\\
 $2^{3}P_{1} \rightarrow 1^{3}D_{1}$  &4.19 &  & 5.35 & 11& & 31.15(21.53)&\\
 $2^{3}P_{0} \rightarrow 1^{3}D_{1}$ &2.31 &  & 3.21 &  1.4& &33.24(13.55)& \\
 \hline\hline
 \end{tabular*}
 %\end{center}
 %\end{ruledtabular}
\label{Table:em1}
 \end{table*}

\begin{table*}[!htb]
\caption{Magnetic dipole (M1) transitions widths. (LP = Linear potential model, SP = Screened potential model, NR = Non-relativistic and  RE = Relativistic, Here $\Gamma$ in KeV) }
%\begin{center}
\begin{tabular*}{\textwidth}{@{\extracolsep{\fill}}cccccccc}
\hline\hline
Transition & Present & \cite{Tanabashi:2018oca} & \cite{Kher:2018wtv} & \cite{Cao:2012du}& \cite{Deng:2016stx}& \cite{Sultan:2014oua}&\cite{Soni:2017wvy}\\
 & work & Expt. &  &  & LP(SP)& RE(NR)&\\
\hline

$1^{3}S_{1}\rightarrow1^{1}S_{0}$  &1.255 &$1.58\pm0.37$ & 1.647 & 2.2 & 2.39 (2.44)& 2.765 (2.752)&1.18\\
$2^{3}S_{1}\rightarrow2^{1}S_{0}$  &1.350 &$0.21\pm0.15$& 0.135  &0.096 &0.19 (0.19) &0.198 (0.197)&0.50\\
$3^{3}S_{1}\rightarrow3^{1}S_{0}$  &1.194 &   &0.082 & 0.044 & 0.051 (0.088)&0.023 (0.044)&0.36\\
$2^{3}S_{1}\rightarrow1^{1}S_{0}$  &0.846 &$1.24\pm0.29$ & 69.57 & 3.8&8.08 (7.80)& 3.370 (4.532)&3.25\\
$2^{1}S_{0}\rightarrow1^{3}S_{1}$  &4.060 &  &35.72 & 6.9&2.64 (2.29)& 5.792 (7.962)& \\
\hline

$1^{3}P_{2}\rightarrow1^{3}P_{0}$  & 8.043&   & 1.638  &&&& \\
$1^{3}P_{2}\rightarrow1^{3}P_{1}$  &1.592 &   &0.189   &&&& \\
$1^{3}P_{2}\rightarrow1^{1}P_{1}$ &0.245 &  & 0.056  &&&& \\
$1^{1}P_{1}\rightarrow1^{3}P_{0}$  &2.768 &  &0.782   &&&& \\
\hline\hline
\end{tabular*}%\end{center}
\label{Table:em2}
\end{table*}

\section{Results and Discussion}
\label{sec:result}
\begin{figure}
\includegraphics[width=0.5\textwidth]{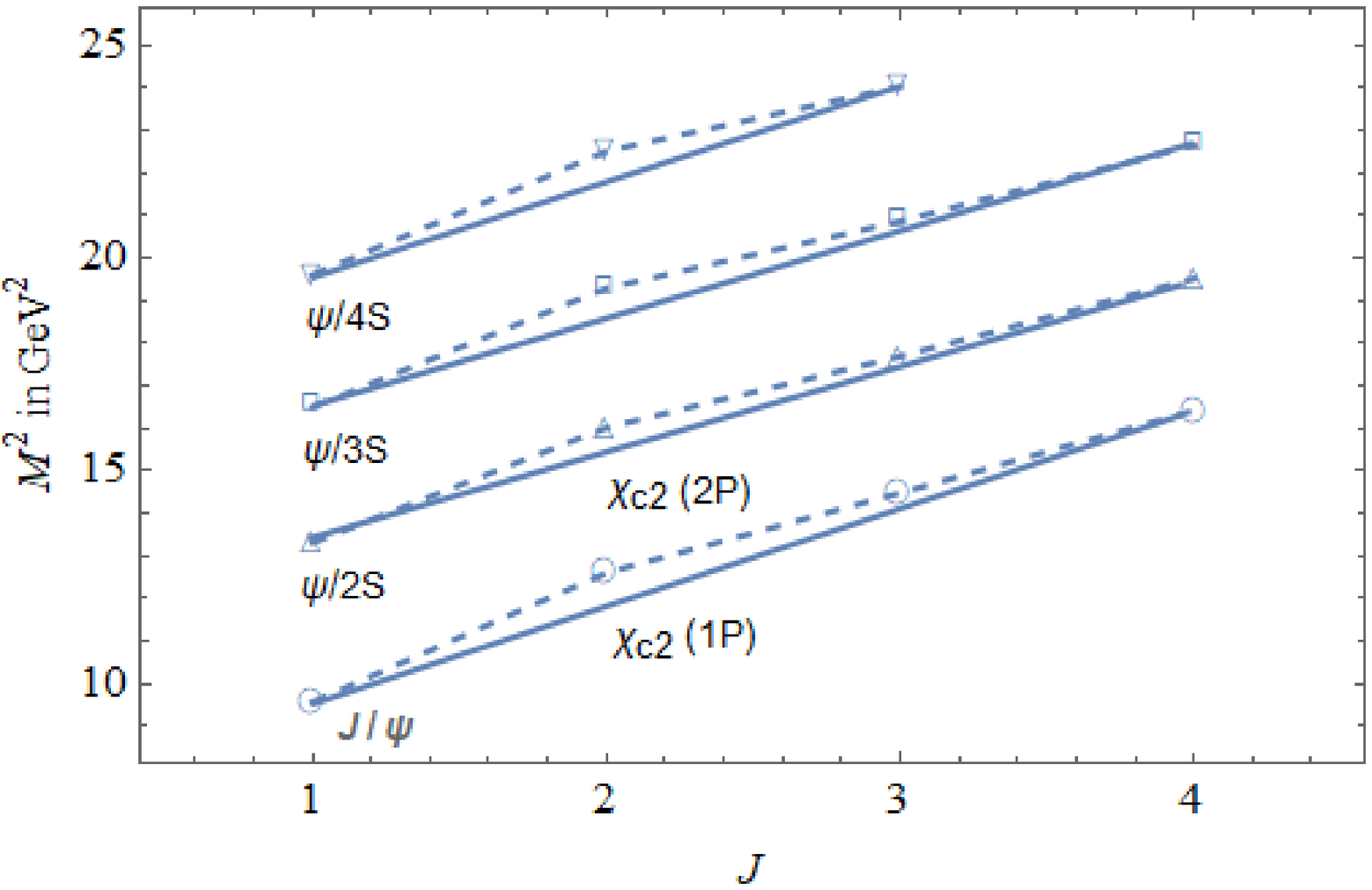}
\caption{$(J,M^2)$ Regge trajectory of parent and daughter for charmonium with un-natural parity. Solid shapes indicates predicted mass, hollow circle represent experimental masses }\label{fig:1}
\end{figure}

\begin{figure}
\includegraphics[width=0.5\textwidth]{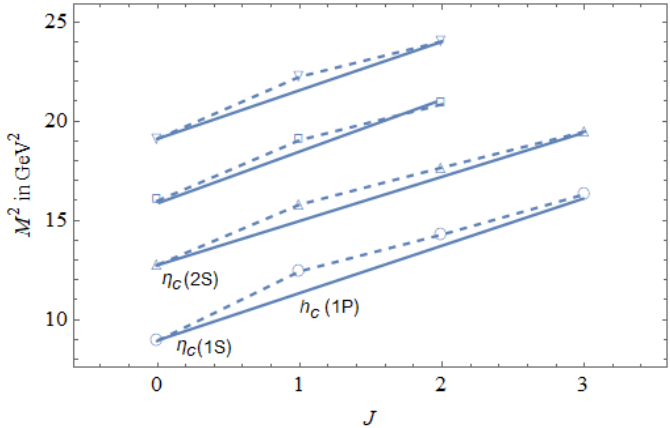}
\caption{Regge trajectory $(J,M^2)$ of $c\bar{c}$ meson with unnatural parity}
\label{fig:2}
\end{figure}

\begin{figure}
\includegraphics[width=0.5\textwidth]{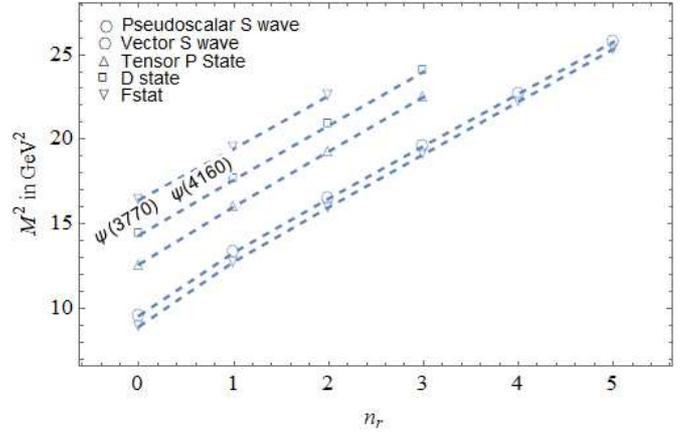}
\caption{Regge trajectory $(n_r,M^2)$ for the Pseudoscaler and vector S state, excited P and D state masses of the $c\bar{c}$ meson}
\label{fig:3}
\end{figure}

\begin{figure}
\includegraphics[width=0.5\textwidth]{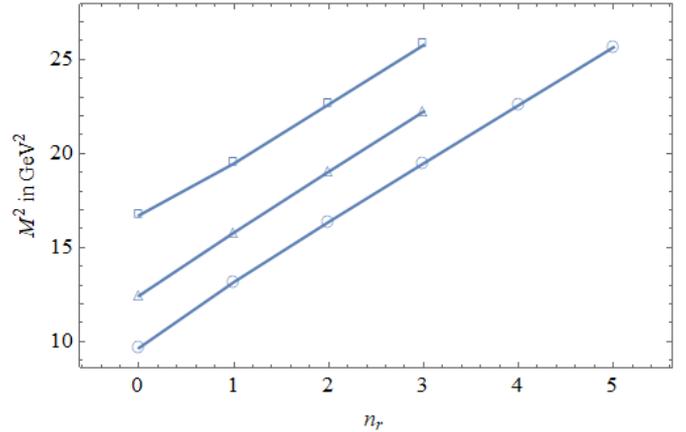}
\caption{Regge trajectory $(n_r,M^2)$  for the S-P-D States center of weight mass for the $c\bar{c}$ meson}
\label{fig:4}
\end{figure}

In order to understand the structure of `XY' Charmoniumlike states, we first calculate complete charmonium mass spectrum by solving the Schr\"{o}dinger equation numerically, spin dependent part of the conventional one gluon exchange potential is employed to obtain mass difference between degenerate mesonic states.
Considering parity constraints, then the Charmoniumlike states are explained as admixtures of pure charmonium states. We compare our calculated mass spectrum(Cornell potential coupled with relativistic correction in the framework of pNRQCD) with experimentally determined masses and also with other theoretical approaches like relativistic model\cite{Ebert:2011jc}, LQCD\cite{Kalinowski:2015bwa}, static potential\cite{Sultan:2014oua} and non-relativistic model i.e considering only the Cornell potential\cite{Chaturvedi:2018xrg,Soni:2017wvy}. We also perform a comparative study of present mass spectra with our previous result\cite{Chaturvedi:2018xrg}. The mass of charm quark and confining strength(A) is same in both the approaches except for the fact that in present approach in addition to confining strength(A) certain other parameters are also incorporated.
Using calculated radial and orbital excited states masses, the Regge trajectories in $(n_r,M^2)$ and $(J,M^2)$ planes are constructed,  with the principal quantum number related to $n_r$ via relation $n_r = n-1$ and $J$ is total angular momentum quantum number. Following equations are used $J = \alpha M^2 + \alpha_0$ and $n_r = \beta M^2 + \beta_0$, where $\alpha, \beta$ are the slopes and $\alpha_0, \beta_0$ are the intercepts. The $(n_r,M^2)$ and $(J,M^2)$ Regge trajectories are plotted in Figures (\ref{fig:1},\ref{fig:2},\ref{fig:3} \& \ref{fig:4}), and slopes and intersepts are tabulated in tables (\ref{Table:reg1},\ref{Table:reg2} \& \ref{Table:reg3}).  Calculated charmonium masses fit well into the linear trajectories in both planes. The trajectories are almost parallel and equidistant and the daughter trajectories appear linear. The Regge trajectories can be helpful for the identification of higher excited state as member of charmonium family.\\

%\begin{figure}
%\includegraphics[width=7cm]{natural.eps}
%\caption{Regge trajectory $(J,M^2)$ of $c\bar{c}$ meson with natural parity}
%\label{fig:1}
%\end{figure}

\begin{table*}[!htb]
\caption{Fitted parameters of the $(J,M^2)$ Regge trajectory with natural and unnatural parity}
%\begin{center}
\begin{tabular*}{\textwidth}{@{\extracolsep{\fill}}cccc}
\hline \hline
Parity&$c\bar c$ & $\alpha (GeV^{-2})$ & $\alpha_0$  \\
\hline
&Parent& 0.439$\pm$0.038&-3.333$\pm$0.515\\
Natural&First Daughter& 0.487$\pm$0.039&-5.598$\pm$0.650\\
&Second Daughter& 0.491$\pm$0.046&-7.230$\pm$0.921\\
&Third Daughter& 0.438$\pm$0.077&-7.660$\pm$1.739\\
\hline
&Parent&0.407$\pm$0.045&-3.786$\pm$0.062\\
Unnatural&First Daughter&0.446$\pm$0.042&-5.828$\pm$0.708\\
&Second Daughter&0.451$\pm$0.046&-7.355$\pm$0.910\\
&Third Daughter&0.398$\pm$0.063&-7.667$\pm$1.380\\
\hline \hline
\end{tabular*}%\end{center}
\label{Table:reg1}
\end{table*}

\begin{table*}[!htb]
\caption{Fitted parameters of Regge trajectory $(n_r,M^2)$ for the S-P-D states}
%\begin{center}
\begin{tabular*}{\textwidth}{@{\extracolsep{\fill}}cccc}
\hline \hline
$c\bar c$& J\^P & $\beta(GeV^{-2})$ & $\beta_0$ \\
\hline
S&$0^{-}$&0.308$\pm$0.006&-2.855$\pm$0.106\\
S&$1^{-}$&0.312$\pm$0.005&-3.082$\pm$0.103\\
P&$2^{+}$&0.303$\pm$0.003&-3.840$\pm$0.051\\
D&$1^{-}$&0.311$\pm$0.002&-4.458$\pm$0.031\\
D&$3^{-}$&0.321$\pm$0.000&-5.270$\pm$0.017\\
\hline \hline
\end{tabular*}%\end{center}
\label{Table:reg2}
\end{table*}

\begin{table*}[!htb]
\caption{ Fitted parameters of Regge trajectory $(n_r,M^2)$ for the S-P-D states center of weight mass}
%\begin{center}
\begin{tabular*}{\textwidth}{@{\extracolsep{\fill}}ccc}
\hline \hline
$c\bar c$  & $\beta(GeV^{-2})$ & $\beta_0$\\
\hline
S&0.314$\pm$0.004&-3.102$\pm$0.068\\
P&0.306$\pm$0.002&-3.823$\pm$0.044\\
D&0.329$\pm$0.007&-5.470$\pm$0.160\\
\hline \hline\\
\end{tabular*}%\end{center}
\label{Table:reg3}
\end{table*}

\subsection{Mass spectra}
The mass difference between the S wave states, $1^1S_0$ - $1^3S_1$ is 105 MeV and $2^1S_0$ - $2^3S_1$ is 114 MeV for the present calculations, while in \cite{Chaturvedi:2018xrg} the mass difference was 82 and 63 MeV respectively, experimentally observed mass difference is 113 and 47 MeV. Thus in the present calculation the splitting in S wave degenerate mesonic states has significantly increased for both 1S and 2S states, bringing the calculated masses nearby the experimental ones. For $3^3S_1$ and $4^3S_1$ states our calculated masses are only 30 MeV and 7 MeV greater than the experimentally observed masses, which is a considerably improved result as compared to\cite{Chaturvedi:2018xrg} and other theoretical studies tabulated. For 1P states it can be seen that the calculate masses of $1^1P_1$ , $1^3P_0$ , $1^3P_1$ and $1^3P_2$ when compared with experimentally observed masses differ only by 02, 59, 04 \& 5 MeV respectively. Based on our calculation of mass spectrum we associate ${{\boldsymbol \chi}_{{c0}}{(3915)}}$ or ${{\mathit X}{(3915)}}$ as $2^3P_0$ state, both of them share same parity and there is no mass difference between our calculated and experimentally observed value. Also, we associate ${{\boldsymbol \chi}_{{c1}}{(3872)}}$\cite{Tanabashi:2018oca} as $2^3P_1$ state because both of them share same parity and as shown in Fig.1 it lies on the curve which inspires us for this association and the mass difference between our calculated and experimental value is 78 MeV. For $2^3P_2$ state the mass difference between our calculated and experimental value is 75 MeV. There are no experimentally observed states for 5S, 6S, 3P and 4P hence we compare our results with theoretical results only and observe that our values are good consonance with the relativistic model\cite{Ebert:2011jc} but suppressed when compared with Cornell potential approach\cite{Chaturvedi:2018xrg,Soni:2017wvy}. There is only one experimentally observed state in D and beyond i.e. $1^3D_2$\cite{Bhardwaj:2013rmw} and our calculated mass is less than experimental value by 22 MeV. We associate ${{\boldsymbol \psi}{(3770)}}$ as $1^3D_1$ state because both of them have same parity and the difference between our calculated and experimentally observed mass is 12 MeV which lies in the error bar. Also, we associate ${{\boldsymbol \psi}{(4160)}}$ as $2^3D_1$ state as both of them have same parity also the mass difference between our calculated and experimental value is only 05 MeV. In $(n_r,M^2)$ Regge trajectory both ${{\boldsymbol \psi}{(3770)}}$ and ${{\boldsymbol \psi}{(4160)}}$ lie on the curve and follow linearity and parallelism thus helping us to associate them with $1^3D_1$ and $2^3D_1$ charmonium states.

\subsection{Decay Properties}
Using the potential parameters and reduced normalized wave function we compute the various decay properties like $\gamma\gamma$, $e^{+}e^{-}$, light hadron and $\gamma\gamma\gamma$ decay widths of various states of charmonium, also the $E1$ \& $M1$ transition widths have been calculated in present study.
The $^1S_0 \rightarrow \gamma\gamma$ decay width within the framework of NRQCD has been calculated by four different approaches, $i.e.$ at NNLO in $\nu$, at NLO in $\nu^2$, at $O(\alpha_s \nu^2)$ and at NLO in $\nu^4$. The calculated results are tabulated in table \ref{Table:gammas} and is compared with experimentally observed decay widths\cite{Tanabashi:2018oca}, relativistic quark model (RQM)\cite{Ebert:2002pp}, heavy quark spin symmetry\cite{Lansberg:2006dw}, relativistic Salpeter model\cite{Kim:2004rz} and the decay widths calculated by conventional Van Royen-Weisskopf formula\cite{Soni:2017wvy,Chaturvedi:2018xrg}. The values of the decay widths calculated at NLO in $\nu^4$ is more convincing then the decay width obtained by the rest three approaches and is nearby the experimental results as well.\\
In addition the $n\chi_{0,2} \rightarrow \gamma\gamma$ decay width by NRQCD mechanism is also calculated at NLO in $\nu^2$ and at NNLO in $\nu^2$, the obtained results are tabulated in table \ref{Table:gammap} and compared with experimental and other theoretical decay widths. It is observed that the calculated decay width by both approaches are more or less same and nearby the experimental decay width.\\
The $^3S_1 \rightarrow e^+ e^-$ decay width within NRQCD framework has been calculated at NLO in $\nu$, NNLO in $\nu$, NLO in $\nu^2$, NLO in $\alpha_s\nu^4$ and NLO in $\alpha_s^2\nu^4$. The results obtained are tabulated in table \ref{Table:ee} and are compared with experimental and other theoretical decay widths. It can be commented that the decay width obtained at NLO in $\alpha_s\nu^4$ is nearby the experimental decay width, as compared with the decay width calculated by other approach.\\
The $^1S_0 \rightarrow LH$ decay width, at NLO in $\nu^2$ and at NNLO in $\nu^2$ has been calculated and the results are tabulated in table \ref{Table:lh}, it can be observed that the calculated decay width from both the current approaches has considerably improved as compared to the previous approach\cite{Chaturvedi:2018xrg} and is same as the experimentally determined decay width.\\
The $^3S_1 \rightarrow \gamma\gamma\gamma$ decay width calculated at NLO $\nu^2$ is tabulated in table \ref{Table:ggg}, the result is found to be in perfect agreement with PDG data \cite{Tanabashi:2018oca}.\\
The results for $E1$ and $M1$ transition for charmonium have been listed in Tables \ref{Table:em1} \& \ref{Table:em2}. The obtained results are compared with experimental and other theoretical approaches like linear potential model, screened potential model, relativistic potential model and non-relativistic potential model. It is observed that for the transitions which are experimentally observed our calculated widths are in excellent agreement. Also for transition widths which are not observed experimentally our calculated values are comparable with values obtained by other theoretical approaches. Ratio of $\frac{\Gamma_{ee}(nS)}{\Gamma_{ee}(1S)}$ in Table \ref{Table:ratio} is consistent with experimental data.
\begin{table*}
\caption{The ratios of $\frac{\Gamma_{e^{+}e^{-}}{{\mathit J / \psi}}(nS)}{\Gamma_{e^{+}e^{-}}{{\mathit J / \psi}{(1S)}}}$ for charmonium states}
%\begin{center}
\begin{tabular*}{\textwidth}{@{\extracolsep{\fill}}cccc}
\hline\hline
$\frac{\Gamma_{e^{+}e^{-}}{{\mathit J / \psi}}(nS)}{\Gamma_{e^{+}e^{-}}{{\mathit J / \psi}{(1S)}}}$&Present&Expt. \cite{Tanabashi:2018oca}& \cite{Bhavsar:2018umj}\\
\hline
$\frac{\Gamma_{e^{+}e^{-}}{{\mathit J / \psi}}(2S)}{\Gamma_{e^{+}e^{-}}{{\mathit J / \psi}{(1S)}}}$&0.295&0.43&0.39\\
\hline
$\frac{\Gamma_{e^{+}e^{-}}{{\mathit J / \psi}}(3S)}{\Gamma_{e^{+}e^{-}}{{\mathit J / \psi}{(1S)}}}$&0.124&--&0.21\\
\hline
$\frac{\Gamma_{e^{+}e^{-}}{{\mathit J / \psi}}(4S)}{\Gamma_{e^{+}e^{-}}{{\mathit J / \psi}{(1S)}}}$&0.067&--&0.11\\
\hline
$\frac{\Gamma_{e^{+}e^{-}}{{\mathit J / \psi}}(5S)}{\Gamma_{e^{+}e^{-}}{{\mathit J / \psi}{(1S)}}}$&0.042&--&0.04\\
\hline\hline
\end{tabular*}%\end{center}
\label{Table:ratio}
\end{table*}

\subsection{Charmoniumlike states as admixtures}
%\subsubsection{Negative parity states}
${{\boldsymbol \psi}{(4230)}}$, ${{\boldsymbol \psi}{(4260)}}$ \& ${{\boldsymbol \psi}{(4360)}}$ all having J$^P$ as $1^-$ and lie within $100$ MeV mass range, all three states can have properties different from conventional charmonia state, detailed literature about them has been discussed in introduction part of this article. Because they are so narrowly placed we explain all of them as admixtures of $3^3S_1$ and $3^3D_1$ charmonia states. As per our calculation all three admixture states have 41\%, 36\% \& 51\% contribution from S wave counterpart, our calculated masses of ${{\boldsymbol \psi}{(4230)}}$, ${{\boldsymbol \psi}{(4260)}}$ \& ${{\boldsymbol \psi}{(4360)}}$ match well with their experimental mass. Except for ${{\boldsymbol \psi}{(4230)}}$ state, calculated masses of ${{\boldsymbol \psi}{(4260)}}$ \& ${{\boldsymbol \psi}{(4360)}}$ lie within error bar suggested by PDG\cite{Tanabashi:2018oca}. Also, to strengthen our claim about these states as admixtures we calculate their leptonic decay widths and compare with experimentally observed result. We observe that our calculated leptonic decay width of ${{\boldsymbol \psi}{(4230)}}$ state when compared with BESIII\cite{Ablikim:2015dlj} is approximately $8$ MeV greater, while leptonic decay widths of ${{\boldsymbol \psi}{(4260)}}$ \& ${{\boldsymbol \psi}{(4360)}}$ when compared with PDG\cite{Tanabashi:2018oca} and BaBar\cite{Lees:2014iua} differs by $1.9$ and $4.4$ MeV respectively. Thus we comment that any of ${{\boldsymbol \psi}{(4230)}}$, ${{\boldsymbol \psi}{(4260)}}$ \& ${{\boldsymbol \psi}{(4360)}}$ states can be admixture of $3^3S_1$ and $3^3D_1$ pure charmonia states.\\
Very less is known about ${{\boldsymbol \psi}{(4390)}}$ both theoretically and experimentally, we in present work try to study it as admixture of $3^3S_1$ and $3^3D_1$, our calculated mass differs from experimental mass by $60$ MeV. Experimental leptonic decay width has not been observed but we calculate it to be $2.11$ eV. Due to lack of experimental evidence and scarce theoretical study we are reluctant to mention it as an admixture state and hope that experiments in future can throw more light on this state.\\
${{\boldsymbol \psi}{(4660)}}$ has been studied as a molecular state and also as admixture state by\cite{Bhavsar:2018umj}. Having J$^P$ as $1^-$ we associate it as admixture state of $4^3S_1$ and $4^3D_1$ having 43\% contribution of $4^3S_1$. Our calculate mass is in perfect agreement with PDG mass, we have also calculated its leptonic decay width, which is approximately $3$ eV less than than decay width observed by Belle\cite{Wang:2014hta}. Based on our study we claim it to be an admixture of $4^3S_1$ and $4^3D_1$ pure charmonium states.

%\subsubsection{Positive parity states}
%Masses of positive parity charged charmoniumlike states namely ${\boldsymbol Z}_{{c}}$'s like ${{\boldsymbol Z}_{{c}}{(3900)}}$, ${{\boldsymbol Z}_{{c}}{(4200)}}$ and ${{\boldsymbol Z}_{{c}}{(4430)}}$ have been calculated by considering them as admixtures of $n^1P_1$ and $n^3P_1$ or $n^1P_1$ and $(n-1)^3P_1$  pure charmonium states. ${{\boldsymbol Z}_{{c}}{(3900)}}$ has been predicted by most theorist as molecule-like structure, we consider it as an admixture of $2^1P_1$ and $1^3P_1$, our calculated mass is significantly suppressed in comparison with the experimentally observed mass, also no comment has been made by us regarding any of its decay property. Hence, we do not comment on it being as an admixture state, more experimental study is required in future on ${{\boldsymbol Z}_{{c}}{(3900)}}$. On ${{\boldsymbol Z}_{{c}}{(4200)}}$ and ${{\boldsymbol Z}_{{c}}{(4430)}}$ there is rarely any theoretical study except for one study which predicts ${{\boldsymbol Z}_{{c}}{(4430)}}$ to be a hadro-charmonium candidate, we based on our calculation of their masses predict them as an admixtures of $2^1P_1$ and $3^3P_1$ and $3^1P_1$ and $4^3P_1$ pure charmonium states. Our calculated masses agree well with the masses predicted by PDG and also lie in the suggested error bar. Thus, we comment ${{\boldsymbol Z}_{{c}}{(4200)}}$ as admixture $2^1P_1$ and $3^3P_1$, and ${{\boldsymbol Z}_{{c}}{(4430)}}$ as admixture of $3^1P_1$ and $4^3P_1$.\\
A detailed description of ${{\boldsymbol \chi}_{{c1}}{(4140)}}$ and ${{\boldsymbol \chi}_{{c1}}{(4274)}}$ both having J$^P$ as $1^+$ has been discussed in introduction, they have been predicted as tetra-quark state or a hybrid state by some theorist, and ${{\boldsymbol \chi}_{{c1}}{(4140)}}$ has been predicted as a pure $3^3P_0$ state by\cite{Bhavsar:2018umj}. Based on our calculation we suggest ${{\boldsymbol \chi}_{{c1}}{(4140)}}$ as $3^3P_1$ and $2^1P_1$ admixture. And we suggest ${{\boldsymbol \chi}_{{c1}}{(4274)}}$ as $3^3P_1$ and $3^1P_1$ admixture. Our calculated mass is in agreement with experimental mass.\\

\section{Conclusion}
\label{sec:conclusion}
After looking at over all mass spectrum and various decay properties we comment that the potential employed here i.e. Cornell potential when coupled with relativistic correction in the framework of pNRQCD is successful in determining mass spectra and decay properties of charmonium.  Thus helping us to support our choice of the potential in explaining quark anti-quark interaction in charmonium. Also, some experimental Charmoniumlike states as an admixture of nearby isoparity states have been explained. The constructed Regge trajectories are helpful for the association of some higher excited states to the family of Charmonium. But from our study we comment that more precise experimental studies is required to associate Charmoniumlike states as pure charmonium or as an exotic, molecular or some other states.

\section{Appendix}

\begin{enumerate}
  \item Explanation of the symbols in Equations \ref{eq:nq1} \& \ref{eq:nq5}. $\psi$ \& $\chi$ are Pauli spinor fields that creates heavy quark and anti-quark, $\overrightarrow{D}$ is gauge covariant spatial derivative, and $\psi^{\dag}$ \& $\chi^{\dag}$ are mixed two fermion operator corresponding to the annihilation (or creation) of $Q\bar{Q}$ pair respectively.

  \item Explanation of various symbols appearing in the short distance coefficients in Equations \ref{eq:nq2}, \ref{eq:nq3}, \ref{eq:nq4}, \ref{eq:p2}, \ref{eq:nq6}, \ref{eq:nq7}, \ref{eq:nq8}, \ref{eq:nq9} \& \ref{eq:nq10}; $Q$ is charge of the charm quark its value is $2/3$, $\alpha$ is electromagnetic running coupling constant its value is $1/137$, $C_f=(N_c^2 -1)/2N_c$ is the Casimir for the fundamental representation, $\alpha_s$ is strong running coupling constant its value is given in table \ref{Table:parameters} and $n_f$ corresponds to the flavour of light quark.

  \item In Equation \ref{eq:nq1} the terms are the operators responsible for $\gamma\gamma$ decay of $n ^1 S_0$ states, where (n=1 to 6). $\left |\left<0|\chi^{\dag}\psi|^1 S_0\right>\right|^2$, $\left
[\left<^1S_0|\psi^{\dag}\chi|0\right>\left<0|\chi^{\dag}\left(-\frac{i}{2}\overrightarrow{D}\right)^2\psi|^1S_0\right>\right]$, \\
$\left
[\left<^1S_0|\psi^{\dag}\left(-\frac{i}{2}\overrightarrow{D}\right)^2\chi|0\right>
\left<0|\chi^{\dag}\left(-\frac{i}{2}\overrightarrow{D}\right)^2\psi|^1S_0\right> \right]$ and $\left[\left<^1S_0|\psi^{\dag}\chi|0\right>\left<0|\chi^{\dag}\left(-\frac{i}{2}\overrightarrow{D}\right)^4\psi|^1S_0\right>\right]$.

\item In Equation \ref{eq:nq5} the terms are the operators responsible for $e^{+}e^{-}$ decay of $n  ^3S_1$ states states, where (n=1 to 6). $\left |\left<0|\chi^{\dag}\psi|^1 S_0\right>\right|^2$,\\ $\left[\left<^3S_1|\psi^{\dag}\sigma\chi|0\right>\left<0|\chi^{\dag}\sigma\left(-\frac{i}{2}\overrightarrow{D}\right)^2\psi|^3S_1\right>\right]$,\\ $\left[\left<0|\chi^{\dag}\sigma\left(-\frac{i}{2}\overrightarrow{D}\right)^2\psi|^3S_1\right> \right]$ and\\
     $\left[\left<^3S_1|\psi^{\dag}\sigma\chi|0\right>
\left<0|\chi^{\dag}\sigma\left(-\frac{i}{2}\overrightarrow{D}\right)^4\psi|^3S_1\right>\right]$.

\item The $F's, G's$ and $H's$ in Equations \ref{eq:nq1} \& \ref{eq:nq5} are expressed in terms of various parameters in Equations \ref{eq:wf1} \& \ref{eq:wf2}.

\item In Equation \ref{eq:me1} the operators are expressed in terms of matrix elements. $\left<^1S_0|{\cal{O}}(^1S_0)|^1S_0\right>$, $\left<^1S_0|{\cal{P}}_1(^1S_0)|^1S_0\right>$ and $\left<^1S_0|{\cal{Q}}^1_1(^1S_0)|^1S_0\right>$ are the matrix elements for the decay of $n ^1 S_0$ states into $\gamma\gamma$. In Equation \ref{eq:me2} the operators $\left<^3S_1|{\cal{O}}(^3S_1)|^3S_1\right>$, $\left<^3S_1|{\cal{P}}_1(^3S_1)|^3S_1\right>$ and\\ $\left<^3S_1|{\cal{Q}}^1_1(^3S_1)|^3S_1\right>$ are the matrix elements for the decay of $n ^3 S_1$ states into $e^{+}e^{-}$.

\item In Equations \ref{eq:wf1} \& \ref{eq:wf2} the matrix elements are expressed in terms of independent non-perturbative regularized and renormalized wave functions at origin. $|R_{P}(0)|^2$ and $|R_{V}(0)|^2$ are the square of the pseudoscalar and vector states wave function.

\item We have computed $\nabla^2R$ term as per \cite{Khan:1995np}
\begin{eqnarray}
 \nabla^2R &=& -x R \frac{M}{2}, r \rightarrow 0
\end{eqnarray}
The binding energy, $x  = M - (2m_Q)$; $M$ is mass of respective mesoni state and $Q$ being charge of the charm quark, $C_F=\frac{4}{3}$ and $\alpha=\frac{1}{137}$.
\end{enumerate}

\bibliographystyle{spphys}
\bibliography{epjc}

\end{document}